\documentclass[twocolumn,floats,nofootinbib,floatfix,aps,pra]{revtex4-2}

\usepackage{amsfonts,amssymb,amsmath}
\usepackage{physics}
\usepackage{color,calc}
\usepackage{bm}
\usepackage{amsbsy,amsmath}
\usepackage{array}
\usepackage{booktabs}
\usepackage{graphicx}
\usepackage{graphics}
\usepackage{eepic,epsfig}

\usepackage{graphicx}
\usepackage{subcaption}

\usepackage{caption}

\usepackage{microtype}

\usepackage[breaklinks,hidelinks,colorlinks=true,linkcolor=blue,citecolor=blue,filecolor=blue,urlcolor=blue]{hyperref}

\usepackage[dvipsnames]{xcolor}
\definecolor{myblue}{RGB}{0, 0, 255}
\usepackage[toc,page]{appendix}

\newcommand{\sinc}{\operatorname{sinc}}

\def\be{ \begin{equation} }
\def\ee{ \end{equation} }
\def\bea{ \begin{eqnarray} }
\def\eea{ \end{eqnarray} }
\def\bse{ \begin{subequations} }
\def\ese{ \end{subequations} }
\def\ba{ \begin{array} }
\def\ea{ \end{array} }
\def\bt{ \begin{tabular} }
\def\et{ \end{tabular} }

\def\to{\rightarrow}

\long\def\/*#1*/{}

\usepackage[utf8]{inputenc}
\usepackage[T1]{fontenc}

\usepackage[english, french]{babel}

\begin{document}

% \title{Theoretical Framework for Helical Deflectors in Time-of-Flight Detection}
%\title{Theory of an RF Helical Deflector in Timing and Time-of-Flight Detection: Modeled as a Capacitor with a Time-Dependent Charge Varying with the Frequency of the Applied Voltage}
% \title{Helical Deflector of Charged Particles: Revisiting Shamaev's Model and From Shamaev’s Model to Theoretical Framework}
% \title{Revisiting Shamaev’s Model of Helical Deflectors: A Path to Theory}
%Revisiting the Shamaev Model and Shamaev Resonance in Radiofrequency Helical Deflector
%\title{A Capacitor Model of the Helical Deflector: Shamaev’s Proposal and the Model in the Book Revisited}
\title{A Capacitor Model of the Helical Deflector: Revisiting Shamaev's Proposal and the Textbook Model}
%https://chatgpt.com/c/6941aefc-f460-832e-bff8-2cc8d30c4924

\author{Hayk L. Gevorgyan\textsuperscript{\hyperref[1]{1},\hyperref[2]{2}}}
\email{hayk.gevorgyan@aanl.am}
\affiliation{\phantomsection\label{1}{\textsuperscript{1}Experimental Physics Division, Alikhanyan National Laboratory (Yerevan Physics Institute), 2 Alikhanyan Brothers St., 0036 Yerevan, Armenia}\\
\phantomsection\label{2}{\textsuperscript{2}Quantum Technologies Division, Alikhanyan National Laboratory (Yerevan Physics Institute), 2 Alikhanyan Brothers St., 0036 Yerevan, Armenia}}

\date{\today}

\begin{abstract}
An RF helical deflector is a type of electron and ion optics device that applies a time-dependent rotating transverse electric or magnetic field by means of sinusoidal RF voltage applied on two opposite conducting helical structures (wires, ribbons or other) to deflect a charged-particle beam into an elliptical trajectory, a bunch into a partial elliptical arc, and a single particle to a point on the ellipse.

It is a perspective indirect timing system being concurrent for reaching picosecond time resolution, and have promise being excellent candidate for high precision time-of-flight detection. As a timing system, it converts the temporal structure, i.e., the arrival time trajectory of an electron beam into a spatial pattern --- particularly, an ellipse in the case of a single-frequency RF voltage and continuous electron pencil beam. 

I propose a \emph{capacitor model} of an RF helical deflector and compare it with the existing \emph{textbook model} \cite{ZhigarevBook, Gevorgian2015}, interpret them and provide understanding of them. Furthermore, I analyze the latter, finding analytical formulas for the applied electric field, ellipse sizes (semi-axes) and rotation angle, lengths of the ellipse line, corresponding to the duration of electron pencil bunches or beams. The present article touches the topics of getting circle on resonance limit, of deflection sensitivity and of time measurement using RMS or average arc lengths.    

\end{abstract}

\maketitle

%keywords: Electron Optics, Ion Beam Optics, Accelerator Physics, Particle Accelerator Physics, Oscilloscope, Timing, Clocks, Classical Physics, Classical Physics Mechanics, Applied Physics, Instrumentation and detectors, Mass Spectrometer, Mass Spectrometers, Time of Flight, Time-of-flight mass spectrometry, Undulators, Resonators, Capacitors, Capacitor  

%%%%%%%%%%%%%%%%%%%%%%%%%%%%%%%%%%%%%%%%%%%%%%%%%%%%%%%%%%%%%%%%%%%%%%%%%%%%%%%%%%%%%%%%%%%%%%%%%%%%%%%%%%%%%%%%%%%%%%%%%%%%%%%%%%%%%%%%%

%()%()%()%()%()%()%()%()%()%()%()%
%()%()%()%()%()%()%()%()%()%()%()%
%()%()%()%()%()%()%()%()%()%()%()%

\section{Introduction\label{Sec:intro}}

%()%()%()%()%()%()%()%()%()%()%()%
%()%()%()%()%()%()%()%()%()%()%()%
%()%()%()%()%()%()%()%()%()%()%()%

Atomic clocks \cite{MajorBook}, while providing the most stable long-term time reference with an internal and absolute standard based on atomic transition frequencies, deliver picosecond-resolved absolute timestamps indirectly. In contrast, RF helical deflectors, as phase-referenced (indirect) timing systems using an RF field as the reference clock, can resolve picosecond-scale relative arrival times. Importantly, there is a huge difference --- RF helical deflectors don’t ``tell time'' in the same way as a clock --- they don’t have absolute timing yet like a clock, but they can resolve relative timing, measuring the relative arrival time of a charged particle within a picosecond time window \cite{Gevorgian2015}.

% Atomic clocks \cite{MajorBook}, despite being the most stable long-term time references, don’t directly operate at picosecond resolution, when RF helical deflectors are. Furthermore, there is a huge difference --- RF helical deflectors don’t ``tell time'' in the same way as a clock --- they don’t have absolute timing like a clock, but they can resolve relative timing, measuring the relative arrival time of a charged particle within a picosecond time window \cite{Gevorgian2015}.

From the other hand the most renowned timing systems are streak cameras, Time-to-Digital Converters (TDCs), RF deflectors, and frequency combs. Streak cameras measure ultrafast photon arrival times from optical, UV, X-ray, or sometimes visible light regions, by means of photocathode, electrostatic deflection plates with a time-varying fast ramp voltage and photoelectron detector. TDCs measure electronic pulse arrival times by digitizing time intervals: combining the coarse timestamp from the fast digital counter driven by a high-frequency electronic clock and the fine timing offset from the delay line/interpolator. In frequency combs, being ``rulers in time'', the event arrival time is measured with respect to known time markers, e.g., phase-coherent comb pulses, phase-locked to a stable RF or microwave clock, which itself can be referenced to an atomic clock.

From the perspective of time-of-flight (TOF) and arrival time detection in electron optics, both streak cameras, RF deflectors, and electrostatic or magnetic TOF systems have an important role. All these systems have one in common --- they use deflection mechanisms and can stand as mass spectrometers (limited to electrons in streak cameras), since the deflection of a charged particle depend on its velocity and mass-to-charge ratio. Deflection mechanisms are electrostatic field in electrostatic TOFs, static magnetic field in magnetic TOFs, electrostatic deflection plates in streak cameras, and RF oscillating electric field in RF deflectors. Two common types of RF deflectors are linear RF deflectors and RF helical deflectors. In the former, two parallel or cylindrical \cite{Ibach1991} electrodes are used to generate an oscillating transverse electric field which is standing in space and linearly polarized, producing a linear deflection in a fixed transverse direction. In the latter, helically wound electrodes (wires or ribbons) create a traveling and rotating transverse RF field along the beam axis, resulting in a non-uniform helical deflection of the charged particles, i.e., deflection in differently oriented angles depending on the arrival time of a charged particle forming an ellipse on the detector screen. In the present article, I show that the helical/circular polarization attributed to this rotating field in \emph{the textbook model} \cite{ZhigarevBook, Gevorgian2015} is only a rough approximation and has a limited validity range.

% Defelction systems: electron microscopes or cathode ray tubes

Metallic helical structure systems are multifunctional being used both as helical undulators \cite{Varfolomeev1980}, helical waveguides \cite{Unger1958,Chu1958,Menachem2011}, helical resonators \cite{Nandi2022,Deng2014,Dyussembayev2022}, helical antennas \cite{DalarssonThesis2015}, helical deflectors \cite{Gevorgian2015, Chernousov2000}, etc. Helical slow-wave structures are widely used as effective broadband transmission lines \cite{Basu1996, Basu1979, Rowe1965, Carter2018}, helical resonators \cite{Corum2001, Martines2021}, helical conductors \cite{Sensiper1951}.

The present article is devoted to the theoretical framework and applications of an RF helical deflector as a timing system, a time-of-flight detector for charged particles, and a mass spectrometer.

The paper is organized as follows. In Sec.~\ref{sec:theorymodel}, I provide a \emph{capacitor model} of the helical deflector and explain \emph{the textbook model} \cite{ZhigarevBook, Gevorgian2015}. Sec.~\ref{sec:ellipse} provides theoretical framework and method for measuring deflection transverse velocities and coordinates, ellipse sizes, its rotation angle. Sec.~\ref{Sec:Comparison} serves as a comparison between \emph{capacitor} and \emph{textbook models}. Sec.~\ref{Sec:Limit} analyzes the limit $\omega \rightarrow \omega_c$, the deflection characteristics, and their comparison in three cases: 1) TOF measured at a screen immediately after the deflector, 2) TOF between the deflector and a screen, assuming that the charged particle leaves the deflector exactly from the axis, 3) total TOF, valid under the paraxial approximation inside the deflector. Here, \emph{screen} refers to any target intercepted by the charged particles, representing a detector --- for example, a multichannel plate. Secs.~\ref{Sec:getcircle},~\ref{Sec:pencilbeam},~\ref{Sec:DS} provide information on the problem of getting circle, the case of \emph{on and off} pencil beam, the deflection sensitivity measures. Finally, Sec.~\ref{Sec:concl} represents comments and conclusions.

\section{A Capacitor Model}\label{sec:theorymodel}
\begin{figure}[t]
\bt{r}
\centerline{\includegraphics[width=0.8\columnwidth]{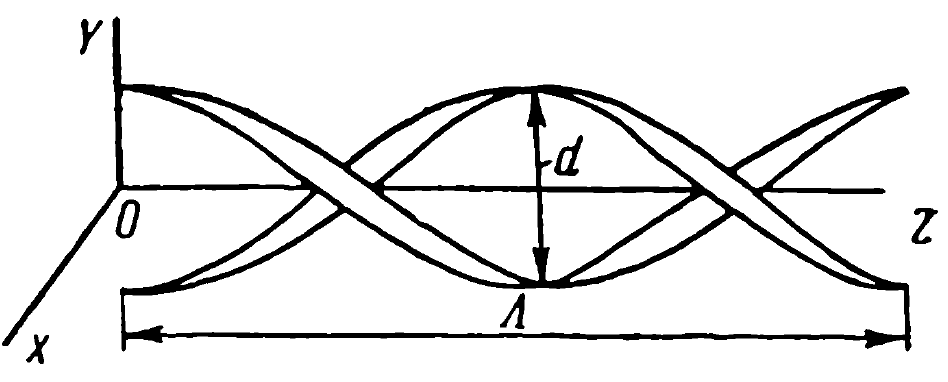}} 
\et
\caption{
Self-unfolding deflection system (helical deflector of charged particles).
}
\label{fig:deflectorpic}
\end{figure}

In 1961, Yu. M. Shamaev proposed a self-unfolding deflection system for the oscillographic observation of UHF (ultrahigh frequency) processes, representing a section of two-wire line wound into a spiral \cite{ZhigarevBook, Gevorgian2015} (see Fig.~\ref{fig:deflectorpic}).

%***************************************************************
%***************************************************************

\subsection{Deflector field}
Equation of helix is
\be\label{helicalline}
x = R \cos{\varphi}, \quad y = R \sin{\varphi}, \quad z = R \kappa \varphi = \frac{\lambda}{2 \pi} \varphi, 
\ee
where $\lambda = l / n$ is a winding pitch of an helix (a solenoidal electrode) --- the axial distance between the two ends of a single turn in a helical winding, where $l$ is the axial distance between the ends of helix, $n$ is the number of turns and can be fractional, $\kappa = \cot{\psi} = \lambda / (2\pi R)$, $\psi$ is a winding pitch angle. When $\varphi$ increases by $2\pi$, $z$ increases by $\lambda$ or $2\pi R \cot{\psi}$, where $R$ is the radius of helix and $\psi$ is the angle of unfolded helix between directions of its axis and linear progression. Note that the length of the helix is $L = l / \cos{\psi} = n \sqrt{\lambda^2 + (2\pi R)^2}$.

The components of the radius vector $\vb*{r}$ and of the infinitesimal element of length $d\vb*{s}$ are
\bse\label{radvec}
\begin{align}
r_x &= x_0 - x = x_0 - R \cos{\varphi}, \quad ds_x = dx = - R \sin{\varphi} d\varphi, \\
r_y &= y_0 - y = y_0 - R \sin{\varphi}, \quad \, \, ds_y = dy = R \cos{\varphi} d\varphi, \\
r_z &= z_0 - z = z_0 - R \kappa \varphi, \quad \quad \, \, \, ds_z = dz = R \kappa d\varphi,
\end{align}
\ese

Using Coulomb's law and assuming that the charge density, arising from a capacitor's negative charge distribution, is uniform $dq (t) = - \rho(t) ds$, the electric field $\vb*{E}(t)$ at a point can be expressed as:
\be\label{Ethelicalline}
\vb*{E}(t) = - \int \frac{\rho(t) \vb*{r}}{4 \pi \epsilon_0 r^3} \, ds,
\ee
where $\rho (t)$ is the time-dependent linear charge density, $\vb*{r}$ is the position vector from the charge element to the observation point, $r = \lvert\vb*{r}\rvert$ is its magnitude, $ds$ represents an infinitesimal element of length, and $\epsilon_0$ is the permittivity of free space. Eqs.~\eqref{helicalline},~\eqref{radvec},~\eqref{Ethelicalline} can be viewed as a \emph{helical line approximation}.

Under the paraxial approximation $x_0 \approx y_0 \approx 0$, the expressions for radius vector $\vb*{r} = r_x \vb*{e}_x + r_y \vb*{e}_y + r_z \vb*{e}_z$, radius $r = \sqrt{r_x^2 + r_y^2 + r_z^2} = \sqrt{R^2 + (z_0 - R \varphi \kappa)^2}$ and length element $ds = \sqrt{ds_x^2 + ds_y^2 + ds_z^2} = R \sqrt{1 + \kappa^2} d\varphi$ from Forms.~\eqref{radvec}, the electric field vector can be written
\be\label{electricfield}
\begin{gathered}
\vb*{E}_\perp (t) = \int\limits_{\varphi_1}^{\varphi_2} Q(t)/R \left\{\begin{array}{c} \cos{\varphi} \\ \sin{\varphi}
\end{array}\right\} \frac{d\varphi}{\left(1 + \left(z_0/R - \kappa \varphi \right)^2 \right)^{3/2}}, \\
E_z = 0.
\end{gathered}
\ee

Since the applied voltage $U(t) = U_0 \sin{(\omega t + \phi)}$ is periodic with a frequency $\omega$ and a phase $\phi$, the linear charge density is also time-tependent given by the same law $\rho (t) = \rho_0 \sin{(\omega t + \phi)}$, hence, $Q(t) = Q_0 \sin{(\omega t + \phi)}$, where $Q_0 = 2 \rho_0 \sqrt{1+ \kappa^2}/(4\pi\epsilon_0)$, where factor of $2$ is taken into account since two helical electrodes create transverse electric field which is twice the transverse electric field generated in a single electrode with capacitance equivalent to linear charge density $\rho (t)$, and $\vb*{E}_\perp (t) = \tilde{\vb*{E}}_\perp (z_0) \sin{(\omega t + \phi)}$, where the formula \eqref{electricfield} is the same for $\tilde{\vb*{E}}_\perp (z_0)$ but $Q_0$ is taken instead of $Q (t)$, and we named $\tilde{\vb*{E}}_\perp (z_0)$ as phasor transverse electric field or phasor with analogy to electromagnetism and optics.

Note that in the non-paraxial theory, the expression~\eqref{electricfield} for the electric field vector changes to 
\be\label{electricfieldnonparax}
\begin{gathered}
\vb*{E} (t) = \int\limits_{\varphi_1}^{\varphi_2} Q(t)/2R \left\{\begin{array}{c} \cos{\varphi} - x_0/R \\ \sin{\varphi} - y_0/R \\ \kappa \varphi - z_0 /R
\end{array}\right\} \frac{d\varphi}{S_1 (x_0, y_0, z_0, \varphi)} + \\
+ \int\limits_{\varphi_1}^{\varphi_2} Q(t)/2R \left\{\begin{array}{c} \cos{\varphi} + x_0/R \\ \sin{\varphi} + y_0/R \\ z_0 /R - \kappa \varphi 
\end{array}\right\} \frac{d\varphi}{S_2 (x_0, y_0, z_0, \varphi)}, \\
S_1 (x_0, y_0, z_0, \varphi) = \left(\left(\frac{x_0}{R} - \cos{\varphi} \right)^2 + \left(\frac{y_0}{R} - \sin{\varphi} \right)^2 + \right. \\ \left. + \left(\frac{z_0}{R} - \kappa \varphi \right)^2 \right)^{3/2}, \\
S_2 (x_0, y_0, z_0, \varphi) = \left(\left(\frac{x_0}{R} + \cos{\varphi} \right)^2 + \left(\frac{y_0}{R} + \sin{\varphi} \right)^2 + \right. \\ \left. + \left(\frac{z_0}{R} - \kappa \varphi \right)^2 \right)^{3/2},
\end{gathered}
\ee
with contributions from the two spirals with the off-axis components of electric fields $\rho_1 (t) = \rho_0 \sin{(\omega t + \phi)}$, $r_{x1} = x_0 - x$, $r_{y1} = y_0 - y$, $r_{z1} = z_0 - z$ and $\rho_2 (t) = - \rho_0 \sin{(\omega t + \phi)}$, $r_{x2} = x_0 + x$, $r_{y2} = y_0 + y$, $r_{z2} = z_0 - z$, where the charged particle's off-axis distance is $\sqrt{x_0^2 + y_0^2}$.

The linear charge density $\rho_0$ and capacitance density (per unit length) $C$ of a two-wire line are given by
\be\label{rho0helicalline}
\rho_0 = C U_0, \quad C = \frac{\pi \epsilon_0 \epsilon_d}{\ln{\frac{d-b}{b}}},
\ee
where $d = 2R$ is the center-to-center distance between the wires, $b$ is the radius of a single wire, $\epsilon_0$ is the permittivity of free space (vacuum), $\epsilon_d$ is the relative permittivity of the dielectric between wires. Eqs.~\eqref{electricfield},~\eqref{electricfieldnonparax},~\eqref{rho0helicalline} are valid in a \emph{helical line approximation}, i.e., when $b \ll d$.

\subsection{Equations of motion}

The equations of motion, using relativistic Lorentz force formula, follows
\be\label{eqmotiongen}
\frac{d \vb*{v}}{dt} = - \frac{e}{m\gamma} \left(\vb*{E}(t) + \vb*{v} \times \vb*{B} (t) \right),
\ee
where $m$ is mass, $e$ is absolute charge, $\gamma = 1/\sqrt{1 - v^2/c^2}$ is Lorentz factor, $v = \sqrt{v_x^2+v_y^2+v_z^2}$ is velocity of an electron; $\vb*{E}(t)$ and $\vb*{B}(t)$ are applied electric and magnetic fields. 

Under the paraxial approximaion ($E_z = 0$ (see Form.~\eqref{electricfield}) and $B_z = 0$ (see Appendix~\ref{Appendix:B})), one gets
\be\label{eqmotion}
\begin{aligned}
\dot{v}_x & \approx - \frac{e}{m\gamma} \left(E_x (t) - v_z B_y (t)\right), \\
\dot{v}_y & \approx - \frac{e}{m\gamma} \left(E_y (t) + v_z B_x (t)\right), \\
\dot{v}_z & = - \frac{e}{m\gamma} \left(v_x B_y (t) - v_y B_x (t) \right) \approx 0, 
\end{aligned}
\ee
where $\gamma \approx 1 / \sqrt{1 - v^2_z/c^2}$. Magnetic field components' contribution of the Lorentz force is much smaller than the electric field contribution (see Appendix~\ref{Appendix:B}). Since the evaluate $[v_z] [B] = \tfrac{c}{10} [B] \ll [E]$ holds, then it follows a fortiori that $[v_x] [B], [v_y] [B] \ll [E]$.

\subsection{\emph{The textbook model} understanding}

Since $z_0 = v_z t$, where $v_z = const$, the phasor's transverse components in the paraxial approximation \eqref{electricfield} can be presented as 
\be\label{theorytomodel}
\begin{gathered}
\tilde{E}_x (z_0) = \tilde{E}_\perp (z_0) \cos{\left(\tilde{\omega}_c (t) t\right)},\\
\tilde{E}_y (z_0) = \tilde{E}_\perp (z_0) \sin{\left(\tilde{\omega}_c (t) t\right)},
\end{gathered}
\ee
where $\tilde{E}_\perp (z_0) = \sqrt{\tilde{E}^2_x (z_0) + \tilde{E}^2_y (z_0)}$ is the magnitude of transverse phasor or just transverse phasor field, which depends on the electron's $z$ coordinate in axis of helix and is rotating, $\tilde{\omega}_c (t) = \arctan{(\tilde{E}_y (t)/\tilde{E}_x (t))} / t$ is time-dependent natural frequency of the system or the frequency of rotation of the field. Thus, the field is frequency- and amplitude-modulated circularly polarized\footnote{Also can be called \emph{helically polarized} due to the field's trajectory when electron is passing through the helical deflector.}, since the frequency is time-dependent (frequency-modulated): $\omega_c (t)$ and the magnitude is space-dependent (amplitude-modulated): $\tilde{E}_\perp (z_0)$.

According to the \emph{textbook model}, the equations of motion inside the helical deflector can be effectively described as 
\be\label{eqmotionShamaev}
\begin{aligned}
\dot{v}_x & \approx - \frac{e}{m\gamma} \tilde{E}_\perp \cos{\left(\omega_c t\right)} \sin{(\omega t +\phi)} \theta(\tau - t) \theta(t), \\
\dot{v}_y & \approx - \frac{e}{m\gamma} \tilde{E}_\perp \sin{\left(\omega_c t\right)} \sin{(\omega t +\phi)} \theta(\tau - t) \theta(t), \\
\dot{v}_z & \approx 0, 
\end{aligned}
\ee

Due to our \emph{capacitor model} and description of the electric field as \eqref{theorytomodel}, we understand the \emph{textbook model} as following. The transverse phasor field is considered with constant magnitude $\tilde{E}_\perp$ and constant frequency $\omega_c = \frac{2\pi v_z}{\lambda}$ (natural frequency of a system), hence is simply circularly polarized. In that case, the equations of motion \eqref{eqmotion} can be solved analytically. In the next section sine ($\sin{x}$) and cosine ($\cos{x}$) functions are denoted as $s_x$ and $c_x$, where $x$ is an argument.

However, the final and approximated values of the field parameters were not specified in his original formulation in the textbook \cite{ZhigarevBook}. We claim that the following approximation holds, i.e, the electric field is approximated by its central value:
\be\label{theoryapprox}
\begin{gathered}
\tilde{E}_\perp (t) \approx \tilde{E}_\perp (t = \tau/2), \\
\tilde{\omega}_c (t) \approx \omega_c,
\end{gathered}
\ee
where $\tau/2$ represents the central value of the temporal range, which is truncated ($[0,\tau]$) using the Heaviside step functions $\theta(x)$. The endpoint $\tau = \frac{R \kappa (\varphi_2 - \varphi_1)}{v_z}$ is derived from the condition $\Delta{z} = R \kappa \Delta{\varphi}$. The use of Heaviside truncation in the \emph{textbook model} is natural, as it defines the temporal bounds of the field.

\begin{figure*}[t]
  \centering
  
  \begin{subfigure}[t]{0.48\textwidth}
    \centering
    \includegraphics[width=\linewidth]{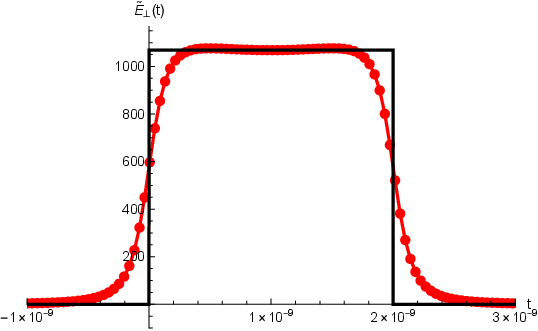}
    \subcaption{$\tilde{E}_\perp (t)$ dependence.}
    \label{fig:Ep_n=1_kappa=6@pi}
  \end{subfigure}
  \hfill
  \begin{subfigure}[t]{0.48\textwidth}
    \centering
    \includegraphics[width=\linewidth]{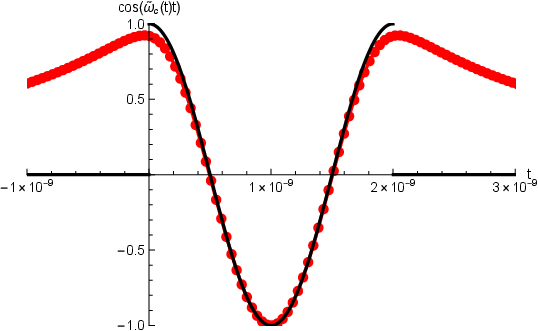}
    \subcaption{$\cos{\left(\tilde{\omega}_c (t) t\right)}$ dependence.}
    \label{fig:cos_n=1_kappa=6@pi}
  \end{subfigure} 
  \vfill
  \begin{subfigure}[t]{0.48\textwidth}
    \centering
    \includegraphics[width=\linewidth]{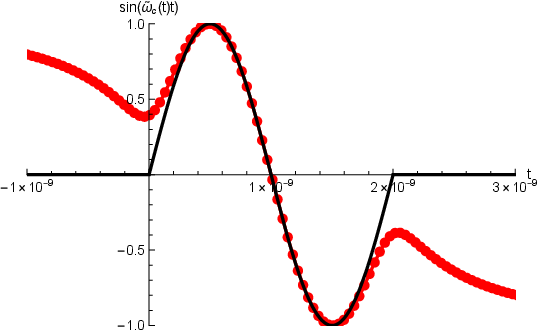}
    \subcaption{$\sin{\left(\tilde{\omega}_c (t) t\right)}$ dependence.}
    \label{fig:sin_n=1_kappa=6@pi}
  \end{subfigure}
  \hfill
  \begin{subfigure}[t]{0.48\textwidth}
    \centering
    \includegraphics[width=\linewidth]{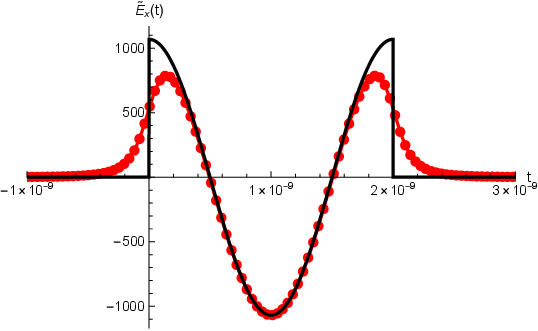}
    \subcaption{$\tilde{E}_x (t)$ dependence.}
    \label{fig:Ex_n=1_kappa=6@pi}
  \end{subfigure}
  \vfill
   \begin{subfigure}[t]{0.48\textwidth}
    \centering
    \includegraphics[width=\linewidth]{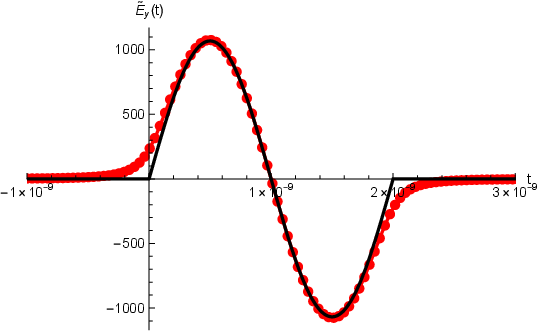}
    \subcaption{$\tilde{E}_y (t)$ dependence.}
    \label{fig:Ey_n=1_kappa=6@pi}
  \end{subfigure}
  
  \caption{\emph{A capacitor model} [\emph{paraxial approximation}] (red) and its \emph{approximation yielding the textbook model} (black) for the case $\kappa = 6/\pi$ and $n=1$ (see Table~\ref{Table:parameters}).}
  \label{FIG:n=1_kappa=6@pi}
\end{figure*}

The \emph{textbook model} is obtained as a central-value approximation of the \emph{capacitor model}, but the \emph{textbook model} uses a Heaviside cutoff, while the \emph{capacitor model} has a continuous field roll-off (see Figs.~\ref{FIG:n=1_kappa=6@pi}-\ref{FIG:kappa=12@pi_n=1}).

We perform a numerical comparison between our \emph{capacitor model} and the corresponding \emph{textbook model} (see, for example, Figs.~\ref{FIG:n=1_kappa=6@pi} and \ref{FIG:n=15_kappa=6@pi}). For large values of $\kappa$ (in our case, already for $\kappa \geq 6/\pi$ or $\lambda/R \geq 12$, see Figs.~\ref{FIG:n=1_kappa=6@pi} and~\ref{FIG:n=15_kappa=6@pi}), the \emph{textbook model} provides a valid approximation, contrary to the case of smaller $\kappa$-s (for example, see the case $\kappa = 1/\pi$ or $\lambda/R = 2$ in Fig.~\ref{FIG:kappa=1@pi_n=1}). Increasing value of $\kappa$ strengthens the validity of approximation (compare Fig.~\ref{FIG:kappa=12@pi_n=1} for $\kappa = 12/\pi$ with Fig.~\ref{FIG:n=1_kappa=6@pi} for $\kappa=6/\pi$) and, contrary, decreasing the value of $\kappa$ weakens it (compare Fig.~\ref{FIG:n=1_kappa=6@pi} for $\kappa = 6/\pi$ with Fig.~\ref{FIG:kappa=1@pi_n=1} for $\kappa=1/\pi$).

\clearpage
\section{Deflector's ellipse sizes on a screen}\label{sec:ellipse}
\subsection{Theory case}\label{subsec:theory}
Integration of equations of motion \eqref{eqmotion} for our theory gives velocities and coordinates in the transverse plane
\be\label{integration}
\begin{aligned}
v_x &= A{'}_1 c_\phi + B{'}_1 s_\phi, \\
v_y &= A{'}_2 c_\phi + B{'}_2 s_\phi, \\
x &= a_1 c_\phi + b_1 s_\phi, \\
y &= a_2 c_\phi + b_2 s_\phi, \\ 
A^\prime_1 &= - \frac{e}{m \gamma} \int\limits_0^\tau \tilde{E}_x (t) s_{\omega t} \, dt ,\\
B^\prime_1 &= - \frac{e}{m \gamma} \int\limits_0^\tau \tilde{E}_x (t) c_{\omega t} \, dt, \\
A^\prime_2 &= - \frac{e}{m \gamma} \int\limits_0^\tau \tilde{E}_y (t) s_{\omega t} \, dt, \\
B^\prime_2 &= - \frac{e}{m \gamma} \int\limits_0^\tau \tilde{E}_y (t) c_{\omega t} \, dt, \\
a_1 &= - \frac{e}{m \gamma} \int\limits_0^\tau \int\limits_0^{\tau^\prime} \tilde{E}_x (t) s_{\omega t} \, dt \, d\tau^\prime , \\
b_1 &= - \frac{e}{m \gamma} \int\limits_0^\tau \int\limits_0^{\tau^\prime} \tilde{E}_x (t) c_{\omega t} \, dt \, d\tau^\prime , \\
a_2 &= - \frac{e}{m \gamma} \int\limits_0^\tau \int\limits_0^{\tau^\prime} \tilde{E}_y (t) s_{\omega t} \, dt \, d\tau^\prime , \\
b_2 &= - \frac{e}{m \gamma} \int\limits_0^\tau \int\limits_0^{\tau^\prime} \tilde{E}_y (t) c_{\omega t} \, dt \, d\tau^\prime .
\end{aligned}
\ee
The coordinates $X(\tau_D, \tau) \overset{\Delta}{=} x(\tau) + X^\prime(\tau_D, \tau)$ and $Y(\tau_D, \tau) \overset{\Delta}{=} y(\tau) + Y^\prime(\tau_D, \tau)$ of an electron on the screen, where $X^\prime(\tau_D, \tau) = v_x (\tau) \tau_D$ and $Y^\prime(\tau_D, \tau) = v_y (\tau) \tau_D$, will be
\be
\begin{aligned}
X = A_1 c_\phi + B_1 s_\phi, \\
Y = A_2 c_\phi + B_2 s_\phi, \\
A_1 = a_1 + A^\prime_1 \tau_D,\\
B_1 = b_1 + B^\prime_1 \tau_D,\\
A_2 = a_2 + A^\prime_2 \tau_D,\\
B_2 = b_2 + B^\prime_2 \tau_D,\\
\end{aligned}
\ee

\begin{figure*}[t]
  \centering
  
  \begin{subfigure}[t]{0.48\textwidth}
    \centering
    \includegraphics[width=\linewidth]{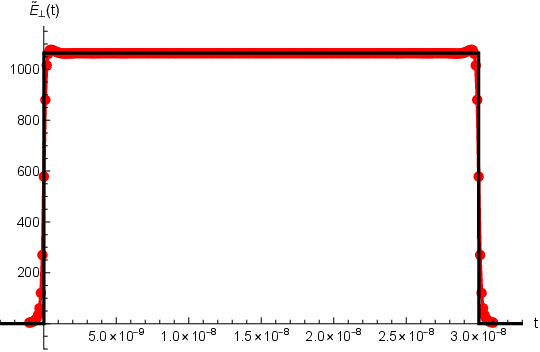}
    \subcaption{$\tilde{E}_\perp (t)$ dependence.}
    \label{fig:Ep_n=15_kappa=6@pi}
  \end{subfigure}
  \hfill
  \begin{subfigure}[t]{0.48\textwidth}
    \centering
    \includegraphics[width=\linewidth]{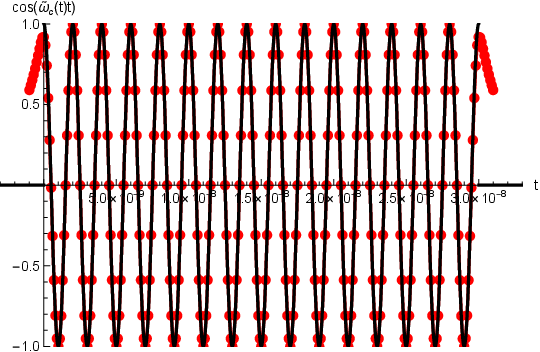}
    \subcaption{$\cos{\left(\tilde{\omega}_c (t) t\right)}$ dependence.}
    \label{fig:cos_n=15_kappa=6@pi}
  \end{subfigure} 
  \vfill
  \begin{subfigure}[t]{0.48\textwidth}
    \centering
    \includegraphics[width=\linewidth]{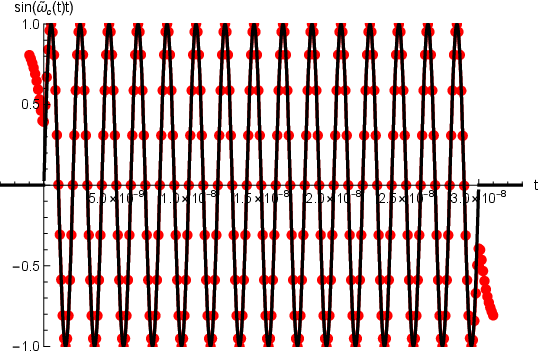}
    \subcaption{$\sin{\left(\tilde{\omega}_c (t) t\right)}$ dependence.}
    \label{fig:sin_n=15_kappa=6@pi}
  \end{subfigure}
  \hfill
  \begin{subfigure}[t]{0.48\textwidth}
    \centering
    \includegraphics[width=\linewidth]{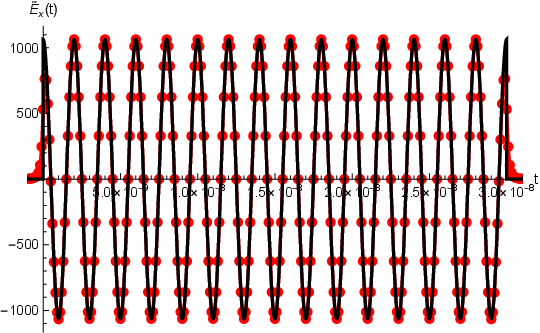}
    \subcaption{$\tilde{E}_x (t)$ dependence.}
    \label{fig:Ex_n=15_kappa=6@pi}
  \end{subfigure}
  \vfill
   \begin{subfigure}[t]{0.48\textwidth}
    \centering
    \includegraphics[width=\linewidth]{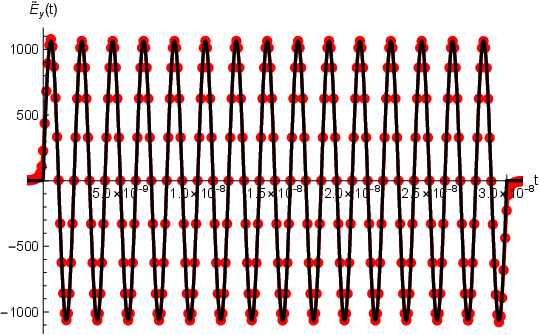}
    \subcaption{$\tilde{E}_y (t)$ dependence.}
    \label{fig:Ey_n=15_kappa=6@pi}
  \end{subfigure}
  
  \caption{\emph{A capacitor model} [\emph{paraxial approximation}] (red) and its \emph{approximation yielding the textbook model} (black) for the case $\kappa = 6/\pi$ (see Table~\ref{Table:parameters}) but $n=15$ ($l=15\times 60$ mm = $90$ cm).}
  \label{FIG:n=15_kappa=6@pi}
\end{figure*} %Top resultsComparison Shamaev and Theory.nb on Desktop (even in overleaf file)

\begin{figure*}[t]
  \centering
  
  \begin{subfigure}[t]{0.48\textwidth}
    \centering
    \includegraphics[width=\linewidth]{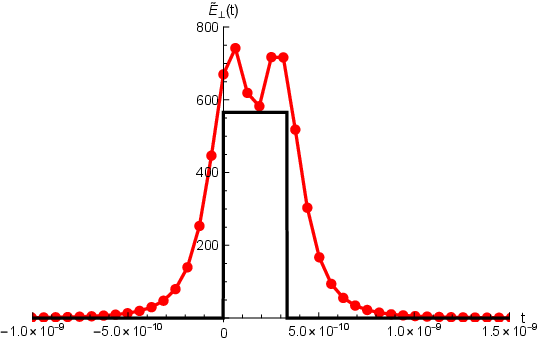}
    \subcaption{$\tilde{E}_\perp (t)$ dependence.}
    \label{fig:Ep_kappa=1@pi_n=1}
  \end{subfigure}
  \hfill
  \begin{subfigure}[t]{0.48\textwidth}
    \centering
    \includegraphics[width=\linewidth]{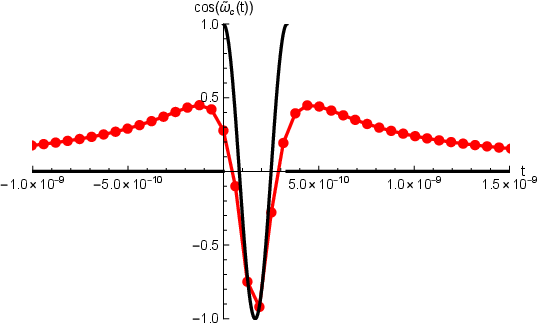}
    \subcaption{$\cos{\left(\tilde{\omega}_c (t) t\right)}$ dependence.}
    \label{fig:cos_kappa=1@pi_n=1}
  \end{subfigure} 
  \vfill
  \begin{subfigure}[t]{0.48\textwidth}
    \centering
    \includegraphics[width=\linewidth]{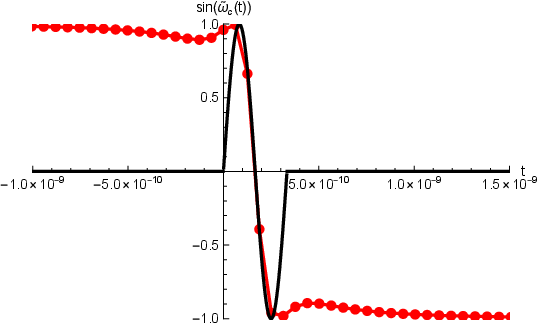}
    \subcaption{$\sin{\left(\tilde{\omega}_c (t) t\right)}$ dependence.}
    \label{fig:sin_kappa=1@pi_n=1}
  \end{subfigure}
  \hfill
  \begin{subfigure}[t]{0.48\textwidth}
    \centering
    \includegraphics[width=\linewidth]{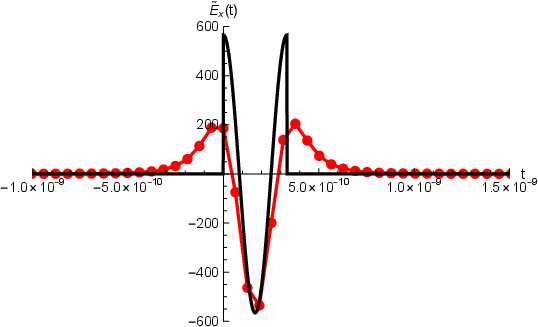}
    \subcaption{$\tilde{E}_x (t)$ dependence.}
    \label{fig:Ex_kappa=1@pi_n=1}
  \end{subfigure}
  \vfill
   \begin{subfigure}[t]{0.48\textwidth}
    \centering
    \includegraphics[width=\linewidth]{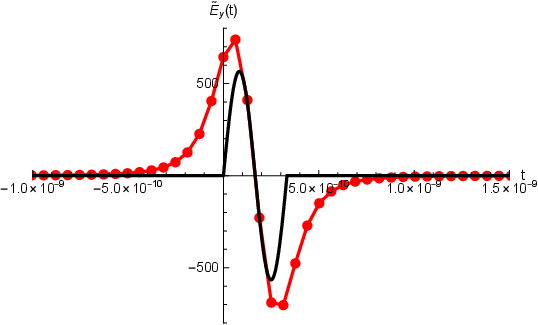}
    \subcaption{$\tilde{E}_y (t)$ dependence.}
    \label{fig:Ey_kappa=1@pi_n=1}
  \end{subfigure}
  
  \caption{\emph{A capacitor model} [\emph{paraxial approximation}] (red) and its \emph{approximation yielding the textbook model} (black) for the case $\kappa = 1/\pi$ ($l=2 R= d = 10$ mm) and $n=1$ (see Table~\ref{Table:parameters}).}
  \label{FIG:kappa=1@pi_n=1}
\end{figure*}

\begin{figure*}[t]
  \centering
  
  \begin{subfigure}[t]{0.48\textwidth}
    \centering
    \includegraphics[width=\linewidth]{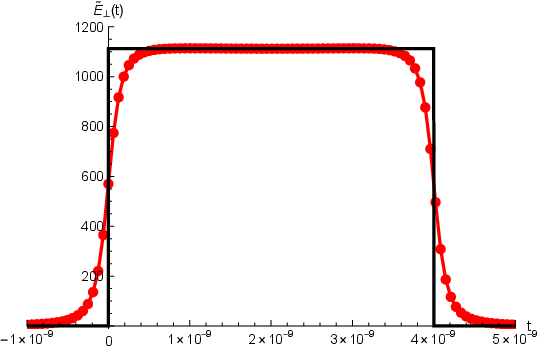}
    \subcaption{$\tilde{E}_\perp (t)$ dependence.}
    \label{fig:Ep_kappa=12@pi_n=1}
  \end{subfigure}
  \hfill
  \begin{subfigure}[t]{0.48\textwidth}
    \centering
    \includegraphics[width=\linewidth]{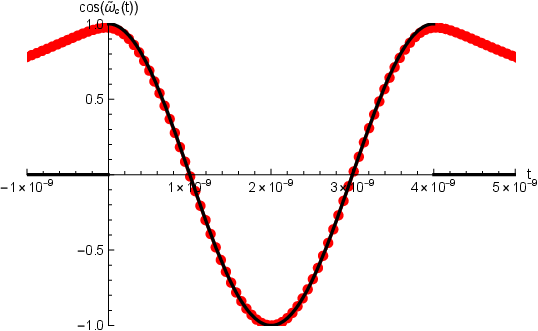}
    \subcaption{$\cos{\left(\tilde{\omega}_c (t) t\right)}$ dependence.}
    \label{fig:cos_kappa=12@pi_n=1}
  \end{subfigure} 
  \vfill
  \begin{subfigure}[t]{0.48\textwidth}
    \centering
    \includegraphics[width=\linewidth]{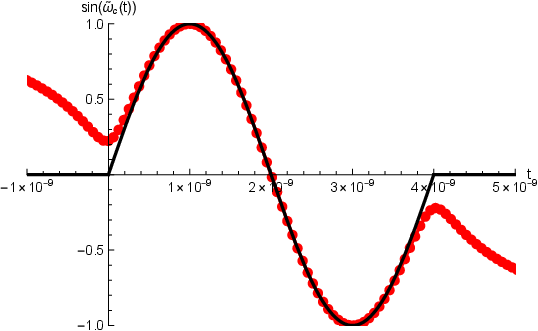}
    \subcaption{$\sin{\left(\tilde{\omega}_c (t) t\right)}$ dependence.}
    \label{fig:sin_kappa=12@pi_n=1}
  \end{subfigure}
  \hfill
  \begin{subfigure}[t]{0.48\textwidth}
    \centering
    \includegraphics[width=\linewidth]{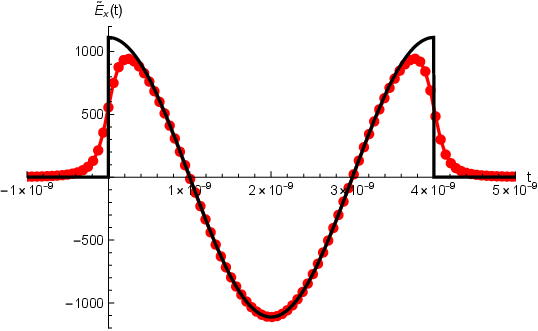}
    \subcaption{$\tilde{E}_x (t)$ dependence.}
    \label{fig:Ex_kappa=12@pi_n=1}
  \end{subfigure}
  \vfill
   \begin{subfigure}[t]{0.48\textwidth}
    \centering
    \includegraphics[width=\linewidth]{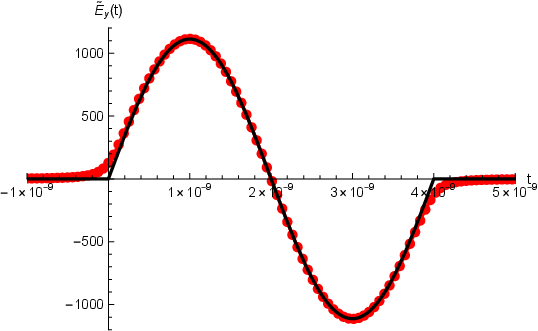}
    \subcaption{$\tilde{E}_y (t)$ dependence.}
    \label{fig:Ey_kappa=12@pi_n=1}
  \end{subfigure}
  
  \caption{\emph{A capacitor model} [\emph{paraxial approximation}] (red) and its \emph{approximation yielding the textbook model} (black) for the case $\kappa = 12/\pi$ ($l=24 R= 12 d = 12$ cm) and $n=1$ (see Table~\ref{Table:parameters}).}
  \label{FIG:kappa=12@pi_n=1}
\end{figure*}

\clearpage
Let's rewrite the equations for coordinates from \eqref{integration} in the form of a single equation of conic sections
\be
\begin{gathered}
a X^2 + b X Y + c Y^2 - 1 = 0, \\
a = \frac{\alpha}{A_1^2} + \frac{\beta}{B_1^2}, \, b = - 2 \left(\frac{\alpha}{A_1 A_2} + \frac{\beta}{B_1 B_2}\right), \, c = \frac{\alpha}{A_2^2} + \frac{\beta}{B_2^2}, \\
\alpha = \frac{1}{\left(\frac{B_1}{A_1} - \frac{B_2}{A_2}\right)^2}, \quad \beta = \frac{1}{\left(\frac{A_1}{B_1} - \frac{A_2}{B_2}\right)^2}.
\end{gathered}
\ee

After doing rotation of the coordinate frame
\be
\begin{bmatrix}
X \\
Y
\end{bmatrix} = \begin{bmatrix}
c_{\theta} & - s_{\theta} \\
s_{\theta} & c_{\theta}
\end{bmatrix} \begin{bmatrix}
x^\prime \\
y^\prime
\end{bmatrix},
\ee
one can find equation $a^\prime (x^\prime)^2 + c^\prime (y^\prime)^2 = 1$ corresponding to the equation of the ellipse in rotated coordinates. Modifying the equation to the standard one $\frac{(x^\prime)^2}{\chi^2} + \frac{(y^\prime)^2}{\varkappa^2} = 1$, one can simply find the rotation angle $\theta$ and the sizes $\chi$ and $\varkappa$ for rotated ellipse
\be
\begin{aligned}
\chi &= \sqrt{\frac{2}{(a-c) 
c_{2\theta} + b s_{2\theta}+a+c}}, \\
\varkappa &= \sqrt{\frac{2}{(c-a) c_{2\theta} - b s_{2\theta} + a + c}}, \\
\chi - \varkappa &= 4 \frac{(a-c) c_{2\theta} + b s_{2\theta}}{(\chi + \varkappa)(a^\prime_0 + b^\prime_0 c_{4\theta} + c^\prime_0 s_{4\theta})}, \\ 
s_{2\theta} & = \frac{b}{\sqrt{(a-c)^2+b^2}}, \\ 
c_{2\theta} & = \frac{a-c}{\sqrt{(a-c)^2+b^2}}, \\
a^\prime_0 &= a^2 - b^2 + c^2 + 6ac, \\
b^\prime_0 &= b^2 - a^2 -c^2 +2 a c,\\
c^\prime_0 &= b c - a b,\\ 
\theta &= \frac{1}{2} \arctan{\left(\frac{b}{a-c}\right)}, \quad \theta \in \left( - \frac{\pi}{2}, \frac{\pi}{2} \right).
\end{aligned}
\ee

Since $\theta$ is an rotation angle, which is necessary to map rotated plane to unrotated one, i.e., after the rotation of the rotated ellipse by this angle we get unrotated ellipse, the rotation angle of the rotated ellipse is actually $-\theta$, because the computational coordinate system is $(x^\prime, y^\prime)$.

Length of the ellipse line from the initial phase $\phi_1$ to the final $\phi_2$ is
\bse\label{lengthinellipse}
\begin{align}
s &= \int\limits_{X(\phi_1)}^{X(\phi_2)} \sqrt{1+\left(\frac{dY}{dX}\right)^2} dx = \\ 
&= \int\limits_{\phi_1}^{\phi_2} \sqrt{\left(\frac{dr(\phi)}{d\phi}\right)^2 + \left(r(\phi)\right)^2} d\phi = \\
&= \int\limits_{\phi_1}^{\phi_2} \sqrt{\left(\frac{dX}{d\phi}\right)^2 + \left(\frac{dY}{d\phi}\right)^2} d\phi = \\
&= \int\limits_{\phi_1}^{\phi_2} \sqrt{A_0 + B_0 c_{2\phi} - C_0 s_{2\phi}} d\phi = \\
&= \int\limits_{\phi_1}^{\phi_2} \sqrt{A_0 - D_0 s_{2\phi-\alpha_0}} d\phi, \\
&= \int\limits_{\phi_1}^{\phi_2} \sqrt{A_0 + D_0 c_{2\phi + \beta_0}} d\phi, \\
&= \sqrt{A_0 + D_0} \int\limits_{\phi_1 + \beta_0 / 2}^{\phi_2 + \beta_0 / 2} 1 - m \sin^2{t} \, dt, \\
&= \sqrt{A_0 + D_0} \left(E[\phi_2 + \beta_0 / 2, m] - E[\phi_1 + \beta_0 / 2, m]\right).
\end{align}
\ese
where $A_0 = \left(A^2_1 + A^2_2 + B^2_1 + B^2_2 \right)/2$, $B_0 = \left(B^2_1 + B^2_2 - A^2_1 - A^2_2\right)/2$, $C_0 = A_1 B_1 +A_2 B_2$, $D_0 = \sqrt{B^2_0 + C^2_0}$, $\alpha_0 = \arctan{(B_0 / C_0)}$, $\beta_0 = \arccot{(B_0/C_0)}$, $m = 2 D_0 / (A_0 + D_0)$, $t = \phi + \beta_0/2$, where the incomplete elliptic integral of the second kind is defined as
\be
E[\theta, m] = \int\limits_{0}^{\theta} 1 - m \sin^2{t} \, dt.
\ee

\subsection{\emph{Textbook model} case}
Integration of equations of motion \eqref{eqmotionShamaev} for the \emph{textbook model}, subject to the initial conditions $v_x (0) = v_y (0) = x (0) = y (0) = 0$, corresponding to an \emph{idealized zero-emittance, space-charge-free} (mutual Coulomb interactions between charges are neglected) or \emph{single-charge pencil beam}, yields the following velocities and coordinates in the transverse plane
\bse\label{velandcoord2}
\begin{align}
v_x (\tau) &= \frac{1}{2} A(\tilde{E}_\perp) \left(\frac{c_\phi}{\omega - \omega_c} + \frac{c_\phi}{\omega + \omega_c} - \right. \notag \\ 
& \left. - \frac{c_{\phi + \tau (\omega - \omega_c)}}{\omega - \omega_c} - \frac{c_{\phi + \tau (\omega + \omega_c)}}{\omega + \omega_c}\right) \\
v_y (\tau) &= \frac{1}{2} A(\tilde{E}_\perp) \left(- \frac{s_{\phi}}{\omega - \omega_c} + \frac{s_{\phi}}{\omega + \omega_c} + \right. \notag \\ 
& \left. + \frac{s_{\phi + \tau (\omega - \omega_c)}}{\omega - \omega_c} - \frac{s_{\phi + \tau (\omega + \omega_c)}}{\omega + \omega_c}\right) \\
x(\tau) &= \frac{1}{2} A(\tilde{E}_\perp) \left(\frac{\tau c_{\phi}}{\omega - \omega_c} + \frac{\tau c_{\phi}}{\omega + \omega_c} + \right. \notag \\ 
& \left. + \frac{s_{\phi}}{(\omega - \omega_c)^2} + \frac{s_{\phi}}{(\omega + \omega_c)^2} - \right. \notag \\ 
& \left. - \frac{s_{\phi + \tau (\omega - \omega_c)}}{(\omega-\omega_c)^2} - \frac{s_{\phi + \tau (\omega + \omega_c)}}{(\omega+\omega_c)^2} \right) \\
y(\tau) &= \frac{1}{2} A(\tilde{E}_\perp) \left(- \frac{\tau s_{\phi}}{\omega - \omega_c} + \frac{\tau s_{\phi}}{\omega + \omega_c} + \right. \notag \\ 
& \left. + \frac{c_{\phi}}{(\omega - \omega_c)^2} - \frac{c_{\phi}}{(\omega + \omega_c)^2} - \right. \notag \\ 
& \left. - \frac{c_{\phi + \tau (\omega - \omega_c)}}{(\omega-\omega_c)^2} + \frac{c_{\phi + \tau (\omega + \omega_c)}}{(\omega+\omega_c)^2} \right),
\end{align}
\ese
where $A(\tilde{E}_\perp) = -\frac{e}{m\gamma} \tilde{E}_\perp$ and $\tau$ is TOF of an electron inside the deflector. Introducing the standard \emph{textbook notations} $x_1 = (\omega_c - \omega) \tau/2$ and $x_2 = (\omega_c + \omega) \tau/2$, the transverse velocities and transverse coordinates can be expressed as  
\begin{widetext}
\bse\label{textbook form}
\begin{align}
v_x (\tau) &= \frac{1}{2} A(\tilde{E}_\perp) \tau \left( - \frac{\sin{x_1}}{x_1} \sin{(x_1 - \phi)} + \frac{\sin{x_2}}{x_2} \sin{(x_2 + \phi)} \right), \\
v_y (\tau) &= \frac{1}{2} A(\tilde{E}_\perp) \tau \left( \frac{\sin{x_1}}{x_1} \cos{(x_1 - \phi)} - \frac{\sin{x_2}}{x_2} \cos{(x_2 + \phi)}\right), \\
x(\tau) &= \frac{1}{4} A(\tilde{E}_\perp) \tau^2 \left( \cos{\phi} \left( - \frac{1}{x_1} \left( 1 - \frac{\sin{2 x_1}}{2 x_1} \right) + \frac{1}{x_2} \left( 1 - \frac{\sin{2 x_2}}{2 x_2} \right) \right) + \sin{\phi} \left( \frac{\sin^2{x_1}}{x^2_1} + \frac{\sin^2{x_2}}{x^2_2} \right) \right), \\
y(\tau) &= \frac{1}{4} A(\tilde{E}_\perp) \tau^2 \left( \cos{\phi} \left( \frac{\sin^2{x_1}}{x^2_1} - \frac{\sin^2{x_2}}{x^2_2} \right) + \sin{\phi} \left( \frac{1}{x_1} \left( 1 - \frac{\sin{2 x_1}}{2 x_1} \right) + \frac{1}{x_2} \left( 1 - \frac{\sin{2 x_2}}{2 x_2} \right)\right) \right)
\end{align}
\ese
\end{widetext}

Transverse coordinates on the screen follow as

\begin{widetext}
\bse
\begin{align}
X(\phi) &= A_1 c_{\phi} + B_1 s_{\phi}, \\
Y(\phi) &= A_2 c_{\phi} + B_2 s_{\phi}, \\
A_1 = \frac{1}{2} \frac{A(\tilde{E}_\perp)}{(\omega^2 - \omega_c^2)^2} & \left(C_0^{A_1} + C_+^{A_1} c_{\tau (\omega + \omega_c)} + S_+^{A_1} s_{\tau (\omega + \omega_c)} + C_-^{A_1} c_{\tau (\omega - \omega_c)} + S_-^{A_1} s_{\tau (\omega - \omega_c)} \right), \\
C_0^{A_1} &= 2 (\tau+ \tau_D) \omega (\omega^2 - \omega^2_c), \notag \\
C_+^{A_1} &= - \tau_D (\omega - \omega_c)^2 (\omega + \omega_c), \notag \\
S_+^{A_1} &= - (\omega - \omega_c)^2, \notag \\
C_-^{A_1} &= -\tau_D (\omega-\omega_c) (\omega+\omega_c)^2, \notag \\
S_-^{A_1} &= - (\omega+\omega_c)^2, \notag \\
B_1 = \frac{1}{2} \frac{A(\tilde{E}_\perp)}{(\omega^2 - \omega_c^2)^2} & \left(C_0^{B_1} + C_+^{B_1} c_{\tau (\omega + \omega_c)} + S_+^{B_1} s_{\tau (\omega + \omega_c)} + C_-^{B_1} c_{\tau (\omega - \omega_c)} + S_-^{B_1} s_{\tau (\omega - \omega_c)} \right), \\
C_0^{B_1} &= 2 (\omega^2 + \omega_c^2), \notag \\
C_+^{B_1} &= - (\omega - \omega_c)^2, \notag \\
S_+^{B_1} &= \tau_D (\omega - \omega_c)^2 (\omega + \omega_c), \notag \\
C_-^{B_1} &= - (\omega + \omega_c)^2, \notag \\
S_-^{B_1} &=  \tau_D (\omega - \omega_c) (\omega + \omega_c)^2 , \notag \\
A_2 = \frac{1}{2} \frac{A(\tilde{E}_\perp)}{(\omega^2 - \omega_c^2)^2} & \left(C_0^{A_2} + C_+^{A_2} c_{\tau (\omega + \omega_c)} + S_+^{A_2} s_{\tau (\omega + \omega_c)} + C_-^{A_2} c_{\tau (\omega - \omega_c)} + S_-^{A_2} s_{\tau (\omega - \omega_c)} \right), \\
C_0^{A_2} &= 4 \omega \omega_c, \notag \\
C_+^{A_2} &= (\omega - \omega_c)^2, \notag \\
S_+^{A_2} &= - \tau_D (\omega - \omega_c)^2 (\omega + \omega_c),\\
C_-^{A_2} &= - (\omega + \omega_c)^2, \notag \\
S_-^{A_2} &= \tau_D (\omega - \omega_c) (\omega + \omega_c)^2, \notag \\
B_2 = \frac{1}{2} \frac{A(\tilde{E}_\perp)}{(\omega^2 - \omega_c^2)^2} & \left(C_0^{B_2} + C_+^{B_2} c_{\tau (\omega + \omega_c)} + S_+^{B_2} s_{\tau (\omega + \omega_c)} + C_-^{B_2} c_{\tau (\omega - \omega_c)} + S_-^{B_2} s_{\tau (\omega - \omega_c)} \right), \\
C_0^{B_2} &= - 2 (\tau + \tau_D) \omega_c (\omega^2 - \omega^2_c), \notag \\
C_+^{B_2} &= - \tau_D (\omega - \omega_c)^2 (\omega + \omega_c), \notag \\
S_+^{B_2} &= - (\omega - \omega_c)^2, \notag \\
C_-^{B_2} &= \tau_D (\omega - \omega_c) (\omega + \omega_c)^2, \notag \\
S_-^{B_2} &= (\omega + \omega_c)^2, \notag 
\end{align}
\ese
\end{widetext}
where $\tau_D$ is TOF of an electron outside the deflector, between the deflector and the screen.

\begin{widetext}
Transverse velocities at the exit of deflector follow  
\bse
\begin{align}
v_x(\phi) &= A^\prime_1 c_{\phi} + B^\prime_1 s_{\phi}, \\
v_y(\phi) &= A^\prime_2 c_{\phi} + B^\prime_2 s_{\phi}, \\
A^\prime_1 = \frac{1}{2} \frac{A(\tilde{E}_\perp)}{(\omega^2 - \omega_c^2)} & \left(F_0 - F_- c_{\tau (\omega + \omega_c)} - F_+ c_{\tau (\omega - \omega_c)} \right), \\
B^\prime_1 = \frac{1}{2} \frac{A(\tilde{E}_\perp)}{(\omega^2 - \omega_c^2)} & \left(F_- s_{\tau (\omega + \omega_c)} + F_+ s_{\tau (\omega - \omega_c)} \right), \\
A^\prime_2 = \frac{1}{2} \frac{A(\tilde{E}_\perp)}{(\omega^2 - \omega_c^2)} & \left(- F_- s_{\tau (\omega + \omega_c)} + F_+ s_{\tau (\omega - \omega_c)} \right), \\
B^\prime_2 = \frac{1}{2} \frac{A(\tilde{E}_\perp)}{(\omega^2 - \omega_c^2)} & \left(- F_c - F_- c_{\tau (\omega + \omega_c)} + F_+ c_{\tau (\omega - \omega_c)} \right), \\
F_0 &= 2 \omega, \notag \\
F_c &= 2 \omega_c, \notag \\
F_- &= \omega - \omega_c, \notag \\
F_+ &= \omega + \omega_c, \notag 
\end{align}
\ese
\end{widetext}

\section{Comparison between models\label{Sec:Comparison}}
%%%%%%%%%%%%%%%%%%%%%%%%%%%%%%%%%%%%%%%%%%%%%%%%%%%%%%%%%%%%%%%%%%%%%%%%%%%%%%%%%%%%%%%%%%%%%%%%%%%%%%%%%%%%%%%%%%%%%%%%%%%%%%%%%%%%%%%%%
%%%%%%%%%%%%%%%%%%%%%%%%%%%%%%%%%%%%%%%%%%%%%%%%%%%%%%%%%%%%%%%%%%%%%%%%%%%%%%%%%%%%%%%%%%%%%%%%%%%%%%%%%%%%%%%%%%%%%%%%%%%%%%%%%%%%%%%%%

\begin{figure*}[t]
  \centering
  \begin{subfigure}[t]{0.24\textwidth}
    \centering
    \includegraphics[width=\linewidth]{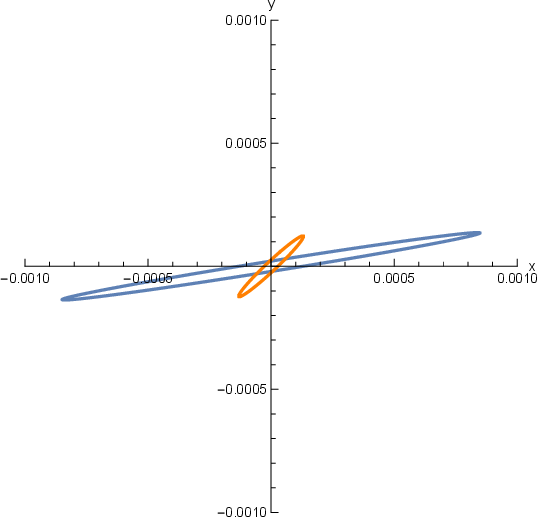}
    \subcaption{$\omega / \omega_c = 0.5$, $n=1/4$.}
    \label{fig:Capacitor_vs_textbook_k=0.5_n=0.25}
  \end{subfigure}
  \hfill
  \begin{subfigure}[t]{0.24\textwidth}
    \centering
    \includegraphics[width=\linewidth]{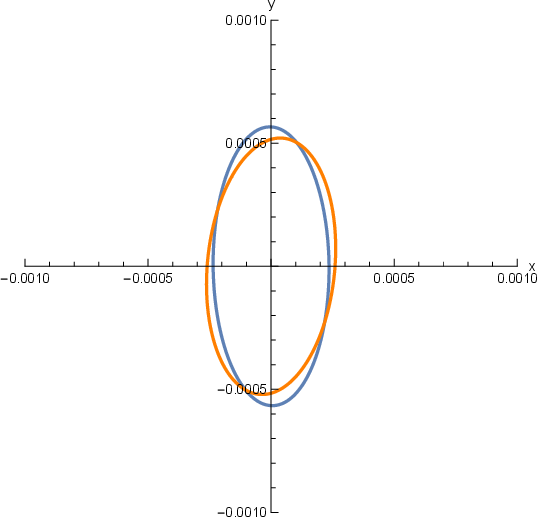}
    \subcaption{$\omega / \omega_c = 0.5$, $n=1/2$.}
    \label{fig:Capacitor_vs_textbook_k=0.5_n=0.5}
  \end{subfigure} 
  \hfill
  \begin{subfigure}[t]{0.24\textwidth}
    \centering
    \includegraphics[width=\linewidth]{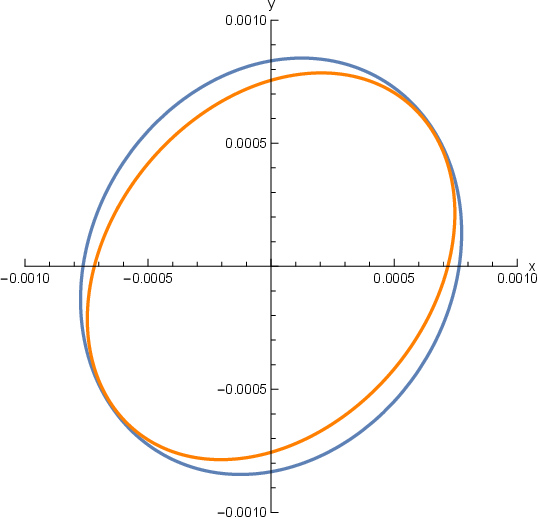}
    \subcaption{$\omega / \omega_c = 0.5$, $n=3/4$.}
    \label{fig:Capacitor_vs_textbook_k=0.5_n=0.75}
  \end{subfigure}
  \hfill
  \begin{subfigure}[t]{0.24\textwidth}
    \centering
    \includegraphics[width=\linewidth]{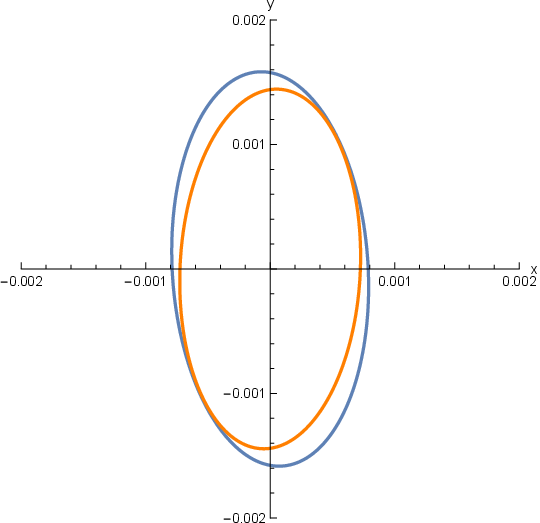}
    \subcaption{$\omega / \omega_c = 0.5$, $n=1$.}
    \label{fig:Capacitor_vs_textbook_k=0.5_n=1}
  \end{subfigure}
  \vfill
  \begin{subfigure}[t]{0.24\textwidth}
    \centering
    \includegraphics[width=\linewidth]{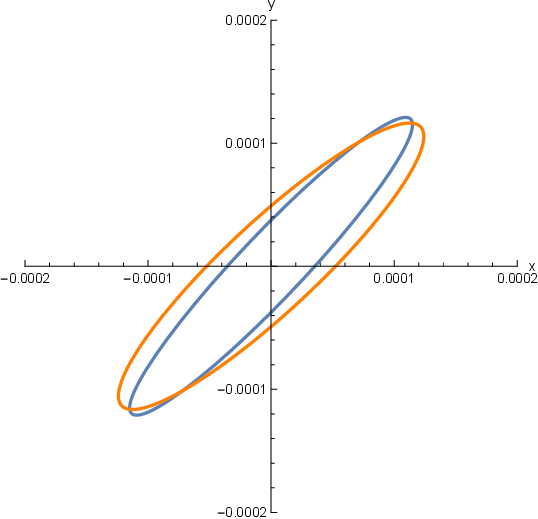}
    \subcaption{$\omega / \omega_c = 1$, $n=1/4$.}
    \label{fig:Capacitor_vs_textbook_k=1_n=0.25}
  \end{subfigure} 
  \hfill
  \begin{subfigure}[t]{0.24\textwidth}
    \centering
    \includegraphics[width=\linewidth]{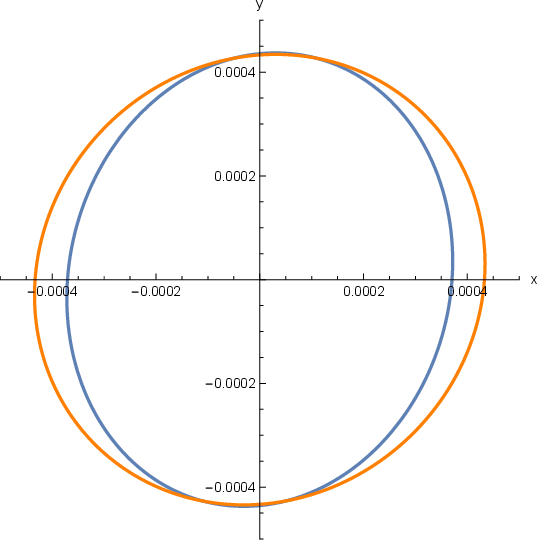}
    \subcaption{$\omega / \omega_c = 1$, $n=1/2$.}
    \label{fig:Capacitor_vs_textbook_k=1_n=0.5}
  \end{subfigure}
  \hfill
  \begin{subfigure}[t]{0.24\textwidth}
    \centering
    \includegraphics[width=\linewidth]{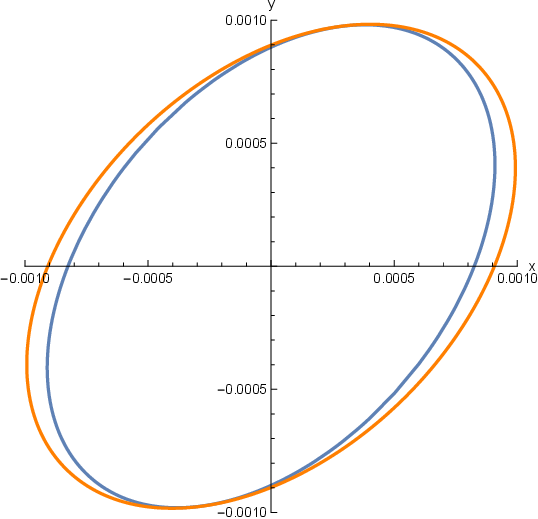}
    \subcaption{$\omega / \omega_c = 1$, $n=3/4$.}
    \label{fig:Capacitor_vs_textbook_k=1_n=0.75}
  \end{subfigure}
  \hfill
  \begin{subfigure}[t]{0.24\textwidth}
    \centering
    \includegraphics[width=\linewidth]{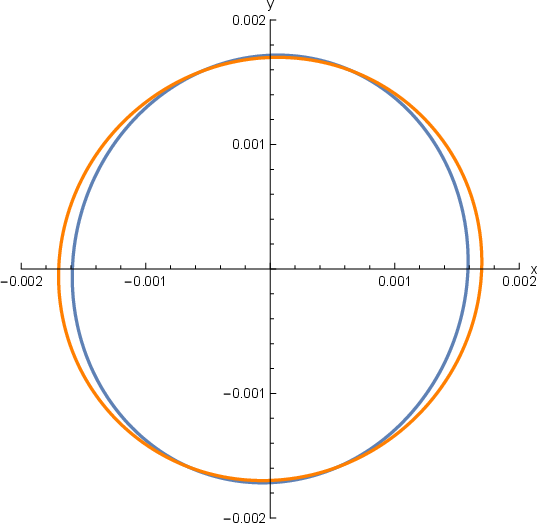}
    \subcaption{$\omega / \omega_c = 1$, $n=1$.}
    \label{fig:Capacitor_vs_textbook_k=1_n=1}
  \end{subfigure}
  \vfill
  \begin{subfigure}[t]{0.24\textwidth}
    \centering
    \includegraphics[width=\linewidth]{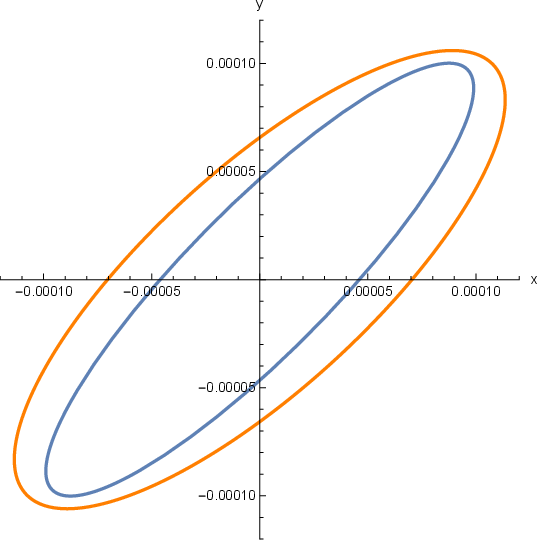}
    \subcaption{$\omega / \omega_c = 1.5$, $n=1/4$.}
    \label{fig:Capacitor_vs_textbook_k=1.5_n=0.25}
  \end{subfigure} 
   \hfill
  \begin{subfigure}[t]{0.24\textwidth}
    \centering
    \includegraphics[width=\linewidth]{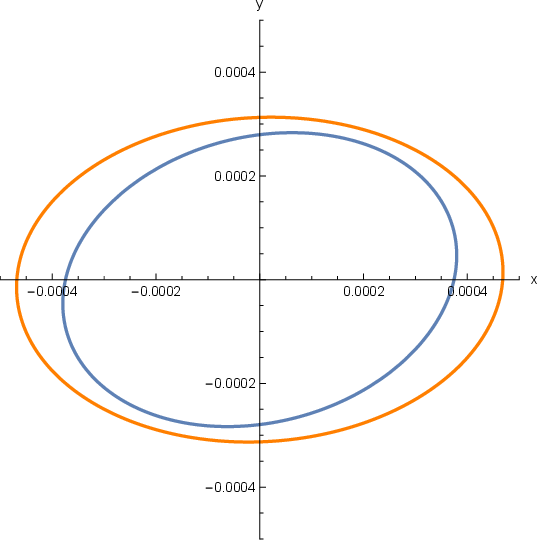}
    \subcaption{$\omega / \omega_c = 1.5$, $n=1/2$.}
    \label{fig:Capacitor_vs_textbook_k=1.5_n=0.5}
  \end{subfigure}
   \hfill
  \begin{subfigure}[t]{0.24\textwidth}
    \centering
    \includegraphics[width=\linewidth]{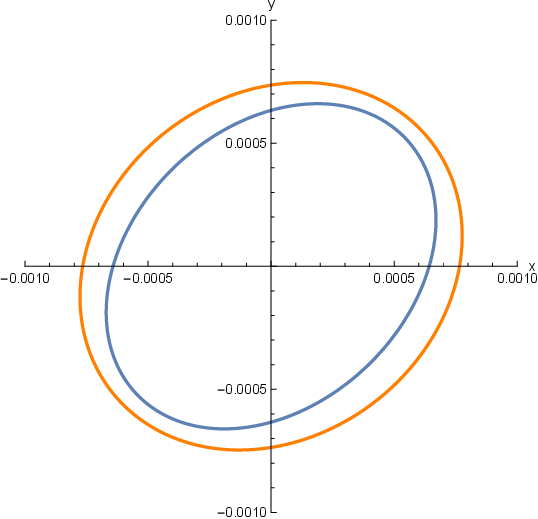}
    \subcaption{$\omega / \omega_c = 1.5$, $n=3/4$.}
    \label{fig:Capacitor_vs_textbook_k=1.5_n=0.75}
  \end{subfigure}  
   \hfill
  \begin{subfigure}[t]{0.24\textwidth}
    \centering
    \includegraphics[width=\linewidth]{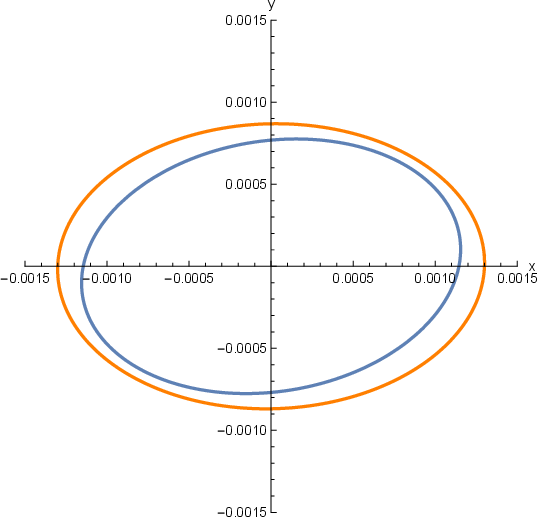}
    \subcaption{$\omega / \omega_c = 1.5$, $n=1$.}
    \label{fig:Capacitor_vs_textbook_k=1.5_n=1}
  \end{subfigure}  
  \vfill
  \begin{subfigure}[t]{0.24\textwidth}
    \centering
    \includegraphics[width=\linewidth]{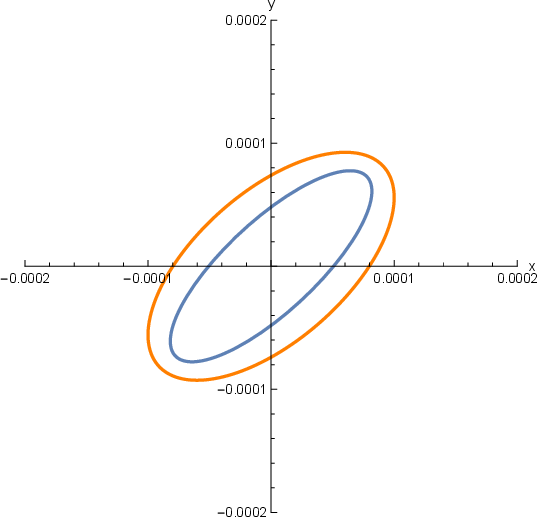}
    \subcaption{$\omega / \omega_c = 2$, $n=1/4$.}
    \label{fig:Capacitor_vs_textbook_k=2_n=0.25}
  \end{subfigure} 
   \hfill
  \begin{subfigure}[t]{0.24\textwidth}
    \centering
    \includegraphics[width=\linewidth]{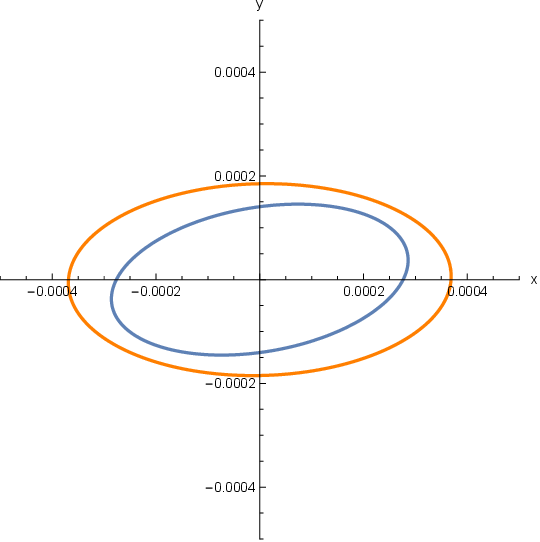}
    \subcaption{$\omega / \omega_c = 2$, $n=1/2$.}
    \label{fig:Capacitor_vs_textbook_k=2_n=0.5}
  \end{subfigure}
   \hfill
  \begin{subfigure}[t]{0.24\textwidth}
    \centering
    \includegraphics[width=\linewidth]{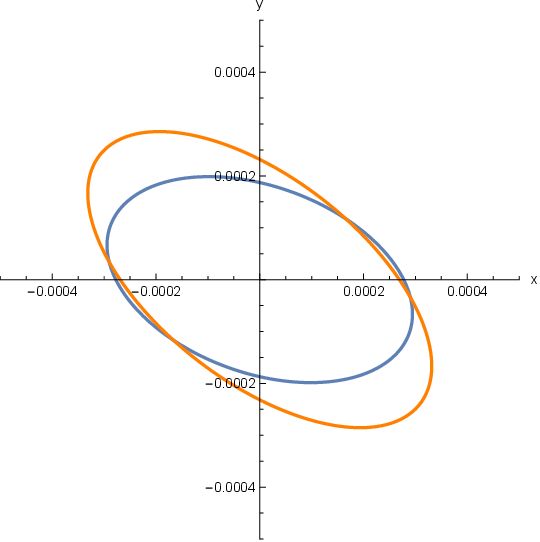}
    \subcaption{$\omega / \omega_c = 2$, $n=3/4$.}
    \label{fig:Capacitor_vs_textbook_k=2_n=0.75}
  \end{subfigure}  
   \hfill
  \begin{subfigure}[t]{0.24\textwidth}
    \centering
    \includegraphics[width=\linewidth]{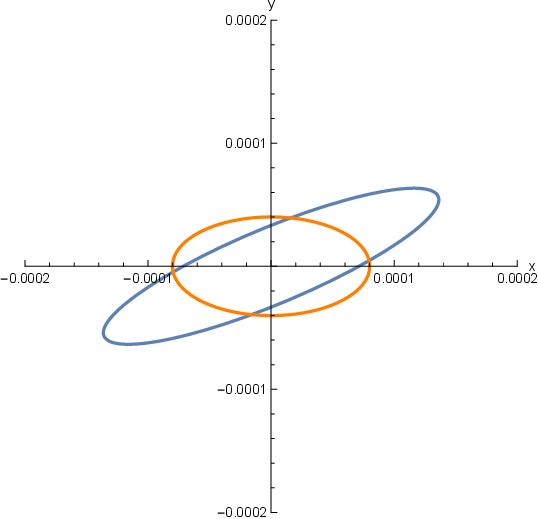}
    \subcaption{$\omega / \omega_c = 2$, $n=1$.}
    \label{fig:Capacitor_vs_textbook_k=2_n=1}
  \end{subfigure} 
   \vfill
  \begin{subfigure}[t]{0.24\textwidth}
    \centering
    \includegraphics[width=\linewidth]{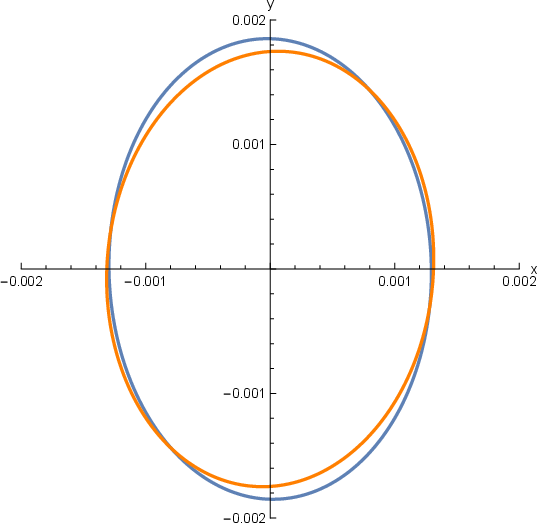}
    \subcaption{$\omega / \omega_c = 0.75$, $n=1$.}
    \label{fig:Capacitor_vs_textbook_k=0.75_n=1}
  \end{subfigure} 
    \hfill
  \begin{subfigure}[t]{0.24\textwidth}
    \centering
    \includegraphics[width=\linewidth]{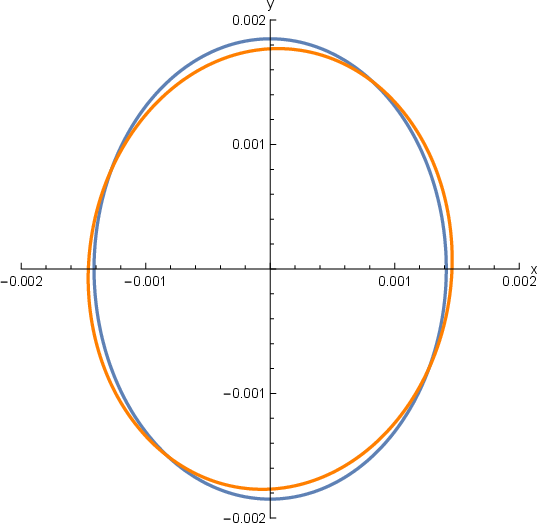}
    \subcaption{$\omega / \omega_c = 0.825$, $n=1$.}
    \label{fig:Capacitor_vs_textbook_k=0.825_n=1}
  \end{subfigure} 
    \hfill
  \begin{subfigure}[t]{0.24\textwidth}
    \centering
    \includegraphics[width=\linewidth]{k=1.5_n=1_Capacitor_paraxial_vs_textbook.eps}
    \subcaption{$\omega / \omega_c = 1.5$, $n=1$.}
    \label{fig:Capacitor_vs_textbook_k=1.5_n=1}
  \end{subfigure} 
  \begin{subfigure}[t]{0.24\textwidth}
    \centering
    \includegraphics[width=\linewidth]{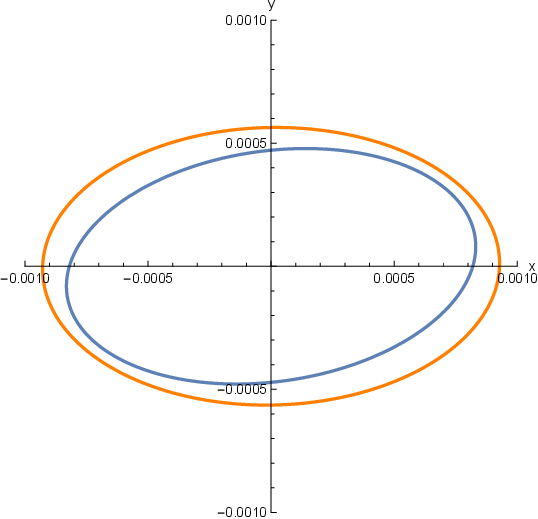}
    \subcaption{$\omega / \omega_c = 1.65$, $n=1$.}
    \label{fig:Capacitor_vs_textbook_k=1.65_n=1}
  \end{subfigure} 
  \caption{\emph{A capacitor model} [\emph{paraxial approximation}] (blue) and \emph{the textbook model} (orange) in the case of Table~\ref{Table:parameters}, zero distance between electron emitter and deflector, and $\tau_D = 4 \tau$ distance between detector screen and deflector.}
  \label{FIG:comparison}
\end{figure*}

Fig.~\ref{FIG:comparison} shows comparison between the resulting ellipses of the two models --- the \emph{capacitor model} in paraxial approximation and the \emph{textbook model} in the case of the parameters in Table~\ref{Table:parameters}, zero distance between electron emitter (electron gun, cathode, electron injector or etc.) and deflector, and $\tau_D = 4 \tau$ distance between detector screen and deflector. Besides, the electric fields' comparison is done between two models in the case of the parameters in Table~\ref{Table:parameters}, $\kappa = 6/\pi$ and $n=1$ (see Fig.~\ref{FIG:n=1_kappa=6@pi}), $\kappa = 6/\pi$ and $n=15$ (see Fig.~\ref{FIG:n=15_kappa=6@pi}), $\kappa = 1/\pi$ and $n=1$ (see Fig.~\ref{FIG:kappa=1@pi_n=1}), $\kappa = 12/\pi$ and $n=1$ (see Fig.~\ref{FIG:kappa=12@pi_n=1}). The differences in ellipses are due to electric field differences at the deflector edges, since the electric field is not zero at the exit of the deflector according to the \emph{capacitor model} (see Figs.~\ref{FIG:n=1_kappa=6@pi}-\ref{FIG:kappa=12@pi_n=1}). Due to these defects from the  \emph{textbook model} at the deflector edges, we have  incorrespondence between the resulting ellipses in two models (see Fig.~\ref{FIG:comparison}), namely, it can be seen in Subfigs.~\ref{fig:Capacitor_vs_textbook_k=0.5_n=0.25} and~\ref{fig:Capacitor_vs_textbook_k=2_n=1}. But in the case of $\omega =\omega_c$ (see Subfigs.~\ref{fig:Capacitor_vs_textbook_k=1_n=0.25}-\ref{fig:Capacitor_vs_textbook_k=1_n=1}), ellipses are approximately equal, and the same for $\omega = 0.5 \omega_c$ the last three Subfigures~\ref{fig:Capacitor_vs_textbook_k=0.5_n=0.5}-\ref{fig:Capacitor_vs_textbook_k=0.5_n=1}, and for $\omega=0.75\omega_c$ and $\omega=0.825\omega_c$ Subfigures~\ref{fig:Capacitor_vs_textbook_k=0.75_n=1}-\ref{fig:Capacitor_vs_textbook_k=0.825_n=1}. In the other cases, when $\omega > \omega_c$, the ellipse for the \emph{textbook model} is larger than for the \emph{capacitor model}. Moreover, the \emph{capacitor model} shows that the case of a circle can not be obtained on resonance $\omega = \omega_c$, since it gives ellipse with minor semiaxis smaller than the circle radius of \emph{textbook model}, when major semiaxis is equal to that radius (see Subfigures~\ref{fig:Capacitor_vs_textbook_k=1_n=0.5} and~\ref{fig:Capacitor_vs_textbook_k=1_n=1}).

%%%%%%%%%%%%%%%%%%%%%%%%%%%%%%%%%%%%%%%%%%%%%%%%%%%%%%%%%%%%%%%%%%%%%%%%%%%%%%%%%%%%%%%%%%%%%%%%%%%%%%%%%%%%%%%%%%%%%%%%%%%%%%%%%%%%%%%%%
%%%%%%%%%%%%%%%%%%%%%%%%%%%%%%%%%%%%%%%%%%%%%%%%%%%%%%%%%%%%%%%%%%%%%%%%%%%%%%%%%%%%%%%%%%%%%%%%%%%%%%%%%%%%%%%%%%%%%%%%%%%%%%%%%%%%%%%%%
\section{Limit\label{Sec:Limit}}
%%%%%%%%%%%%%%%%%%%%%%%%%%%%%%%%%%%%%%%%%%%%%%%%%%%%%%%%%%%%%%%%%%%%%%%%%%%%%%%%%%%%%%%%%%%%%%%%%%%%%%%%%%%%%%%%%%%%%%%%%%%%%%%%%%%%%%%%%
%%%%%%%%%%%%%%%%%%%%%%%%%%%%%%%%%%%%%%%%%%%%%%%%%%%%%%%%%%%%%%%%%%%%%%%%%%%%%%%%%%%%%%%%%%%%%%%%%%%%%%%%%%%%%%%%%%%%%%%%%%%%%%%%%%%%%%%%%
In the resonance limit, $\omega \rightarrow \omega_c$, the TOF required for a charged particle to complete one turn ($n=1$) is equal to the period of the applied RF voltage. For the calculation of the deflection amplitudes and the rotation angle on the resonance $\omega \rightarrow \omega_c$, the following notation will be used
\bse
\begin{gather}
a_c = 2 \tau \omega_c = 4 \pi n, \\
a_D = 2 \tau_D \omega_c = 4 \pi n_D,
\end{gather}
\ese
where the deflector parameter $n = \frac{l}{\lambda}$ is the number of turns of a helix, $n_D = \frac{L_D}{\lambda}$ is the number of \emph{phantom} turns of a helix imagined between the end of the deflector and the screen, and $L_D$ is a distance between deflector and detector screen.

%%%%%%%%%%%%%%%%%%%%%%%%%%%%%%%%%%%%%%%%%%%%%%%%%%%%%%%%%%%%%%%%%%%%%%%%%%%%%%%%%%%%%%%%%%%%%%%%%%%%%%%%%%%%%%%%%%%%%%%%%%%%%%%%%%%%%%%%%
%%%%%%%%%%%%%%%%%%%%%%%%%%%%%%%%%%%%%%%%%%%%%%%%%%%%%%%%%%%%%%%%%%%%%%%%%%%%%%%%%%%%%%%%%%%%%%%%%%%%%%%%%%%%%%%%%%%%%%%%%%%%%%%%%%%%%%%%%
\subsection{TOF in a deflector: on resonance\label{Subsec:TOFdef}}
%%%%%%%%%%%%%%%%%%%%%%%%%%%%%%%%%%%%%%%%%%%%%%%%%%%%%%%%%%%%%%%%%%%%%%%%%%%%%%%%%%%%%%%%%%%%%%%%%%%%%%%%%%%%%%%%%%%%%%%%%%%%%%%%%%%%%%%%%
%%%%%%%%%%%%%%%%%%%%%%%%%%%%%%%%%%%%%%%%%%%%%%%%%%%%%%%%%%%%%%%%%%%%%%%%%%%%%%%%%%%%%%%%%%%%%%%%%%%%%%%%%%%%%%%%%%%%%%%%%%%%%%%%%%%%%%%%%

In the limit $\omega \rightarrow \omega_c \Rightarrow \{x \rightarrow x_c, y \rightarrow y_c\}$ 
\be
\begin{gathered}
x_c = a_{1c} c_\phi + b_{1c} s_\phi, \\ %= a_{xc} \cos{\left(\phi + \phi_{xc}\right)}
y_c = a_{2c} c_\phi + b_{2c} s_\phi, \\ % = a_{yc} \sin{\left(\phi + \phi_{yc}\right)} 
a_{1c} = b_{2c} = \frac{A(\tilde{E}_\perp)}{8 \omega^2_c} \left(a_c - \sin{a_c}\right), \\
b_{1c} = \frac{A(\tilde{E}_\perp)}{8 \omega^2_c} \left(\frac{a^2_c}{2} + 1 - \cos{a_c}\right), \\
a_{2c} = \frac{A(\tilde{E}_\perp)}{8 \omega^2_c} \left(\frac{a^2_c}{2} - (1 - \cos{a_c})\right).
\end{gathered}
\ee

The sizes of the small ellipse are the following
\be
\begin{gathered}
\tilde{\chi}_{c} = \frac{\abs{A(\tilde{E}_\perp)}}{16 \omega^2_c} \left(a^2_c - 2 \sqrt{g(a_c)} \right) = \frac{l^2 \abs{A(\tilde{E}_\perp)}}{v^2_z} \tilde{G}_1 (n), \\
\tilde{\varkappa}_{c} = \frac{\abs{A(\tilde{E}_\perp)}}{16 \omega^2_c} \left(a^2_c + 2 \sqrt{g(a_c)} \right) = \frac{l^2 \abs{A(\tilde{E}_\perp)}}{v^2_z} \tilde{G}_2 (n), \\
\tilde{G}_1 (n) = \frac{1}{4} \left(1 - \frac{\sqrt{g(n)}}{8 \pi^2 n^2} \right), \\
\tilde{G}_2 (n) = \frac{1}{4} \left(1 + \frac{\sqrt{g(n)}}{8 \pi^2 n^2} \right), \\
g(a_c) = 2 + a^2_c - 2 \cos{a_c} - 2 a_c \sin{a_c}, \\
g(n) = 2 + 16 \pi^2 n^2 - 2 \cos(4 \pi n) - 8 \pi n \sin(4 \pi n), 
\end{gathered}
\ee 
where $a^2_c - 2 \sqrt{g(a_c)} \geq 0$, since the function $h(a_c) = a^4_c - 4 g(a_c) \geq 0$, which can be proved by the change of the variable $u = a_c/2$ and the study of its behavior $h(u) = 16 \left(u^4 - (u - \sin{u} \cos{u})^2 - \sin^4{u} \right) \geq 0$. 

%***************************************************************
\begin{figure}[t]
\bt{r}
\centerline{\includegraphics[width=1\columnwidth]{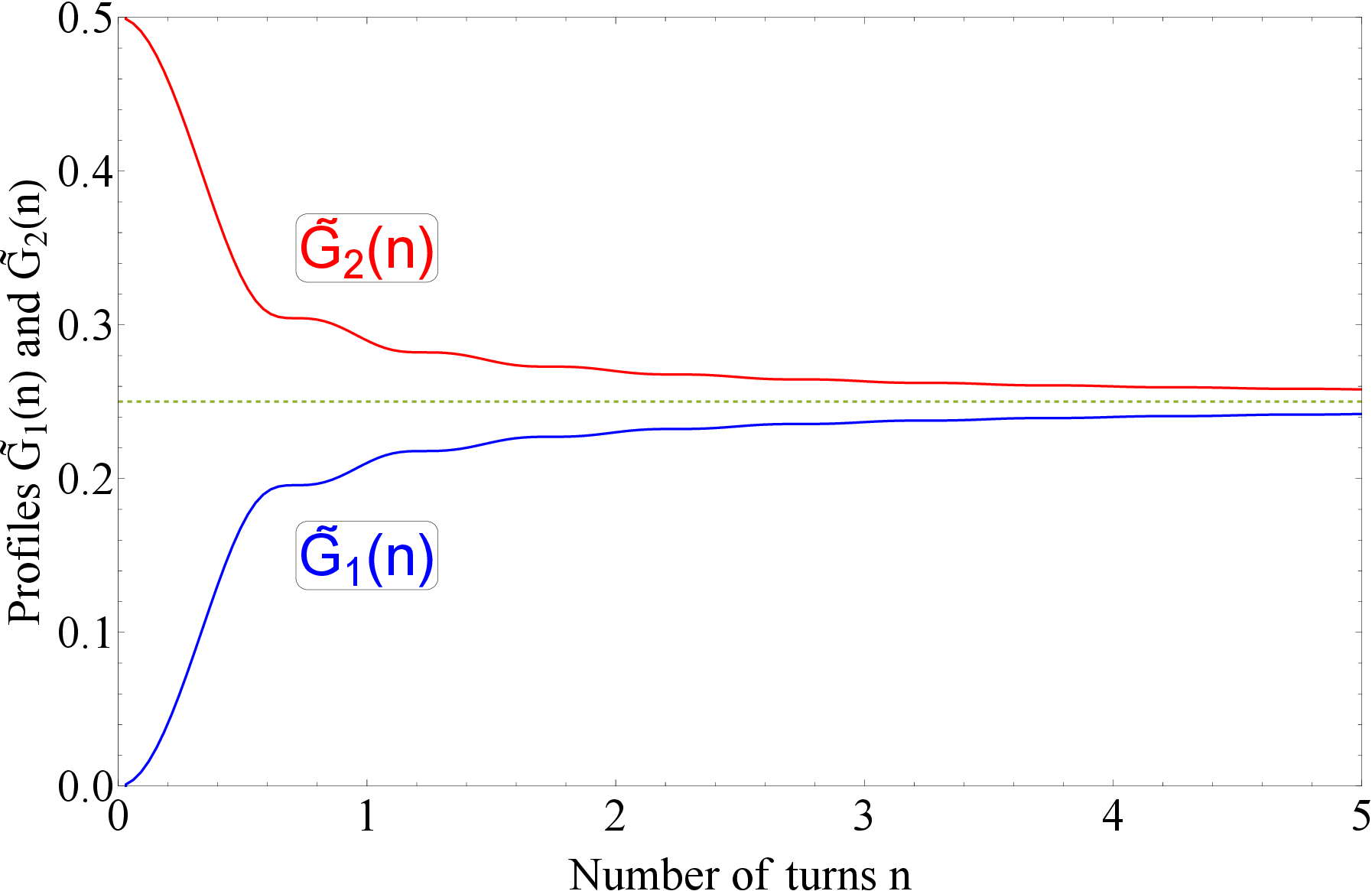}} 
\et
\caption{
$\tilde{G}_1(n)$ and $\tilde{G}_2 (n)$ profile dependences on the number of turns $n$ (the small ellipse on resonance).
}
\label{fig:G1G2ntilde}
\end{figure}
%***************************************************************

$\tilde{G}_1 (n)$ and $\tilde{G}_2 (n)$ profiles from the number of turns are monotonically increasing and decreasing functions (see Fig.~\ref{fig:G1G2ntilde}), respectively: 
\begin{enumerate}
\item for tiny number of turns, $\tilde{G}_1 (n)$ profile has its minimum and $\tilde{G}_2 (n)$ its maximum:
\be
\begin{gathered}
\tilde{G}_{1, \text{min}}(n) = \lim\limits_{n \to 0} \tilde{G}_1 (n) = 0, \\
\tilde{G}_{2, \text{max}}(n) = \lim\limits_{n \to 0} \tilde{G}_2 (n) = \frac{1}{2},
\end{gathered}
\ee
\item when for large number of turns they get closer ($\tilde{G}_{1, \text{max}}(n) = \tilde{G}_{2, \text{max}}(n) = 1/4$):
\be
\begin{gathered}
\tilde{G}_{1, \text{max}}(n) = \lim\limits_{n \to \infty} \tilde{G}_1(n) = \frac{1}{4}, \\
\tilde{G}_{2, \text{min}}(n) = \lim\limits_{n \to \infty} \tilde{G}_2(n) = \frac{1}{4},
\end{gathered}
\ee
and that case corresponds to the case when the ellipse gets closer to the circle $\tilde{\chi}_{c} \rightarrow \tilde{\varkappa}_{c}$ with a radius:
\be
\tilde{r}_{c} = \frac{l^2 \abs{A(\tilde{E}_\perp)}}{4 v^2_z}.
\ee
\end{enumerate}

The \emph{paraxial approximation} holds when $\tilde{\varkappa}_{c} \ll R$ ($\sqrt{x^2(\phi) + y^2(\phi)} \ll d/2$):
\be\label{paraxial}
\begin{gathered}
\frac{e \tilde{E}_\perp}{m \gamma} \frac{l^2}{v^2_z} \tilde{G}_2 (n) \ll R, \\
\end{gathered}
\ee
therefore, the \emph{textbook model} and consequences (derived formulas) are true only when \eqref{paraxial} holds. So, to get a circle, experimentalists must simultaneously ensure that approximations in Sec.~\ref{sec:theorymodel}, condition \eqref{paraxial}, largeness of $n$, and Shamaev conditions hold.

From the validity of the \emph{approximation yielding the textbook model} ($\lambda/R \geq 12$) [see \eqref{theoryapprox} and \eqref{eqmotionShamaev}] and the \emph{paraxial approximation} of the \emph{capacitor model} \eqref{paraxial}, one gets the following two competitive conditions
\be
12 l n \leq \frac{l^2}{R} \ll \frac{v^2_z}{\tilde{G}_2 (n)} \frac{m \gamma}{e \tilde{E}_\perp}.
\ee

%***************************************************************
\begin{figure}[t]
\bt{r}
\centerline{\includegraphics[width=1\columnwidth]{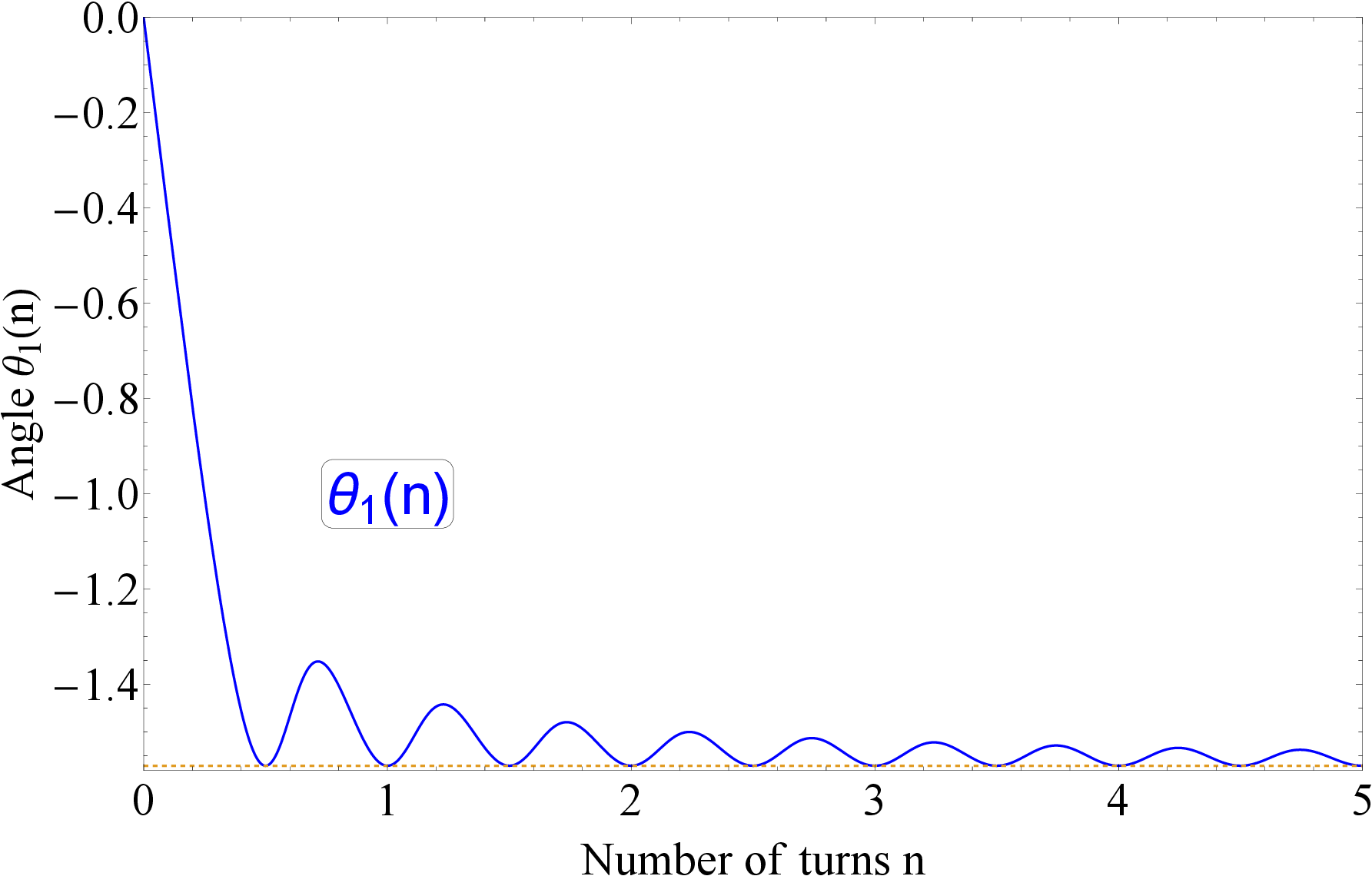}} 
\et
\caption{
Angle $\theta_1(n)$ dependence on the number of turns $n$ (the small ellipse on resonance).
}
\label{fig:theta1n}
\end{figure}
%***************************************************************
%***************************************************************
\begin{figure}[t]
\bt{r}
\centerline{\includegraphics[width=1\columnwidth]{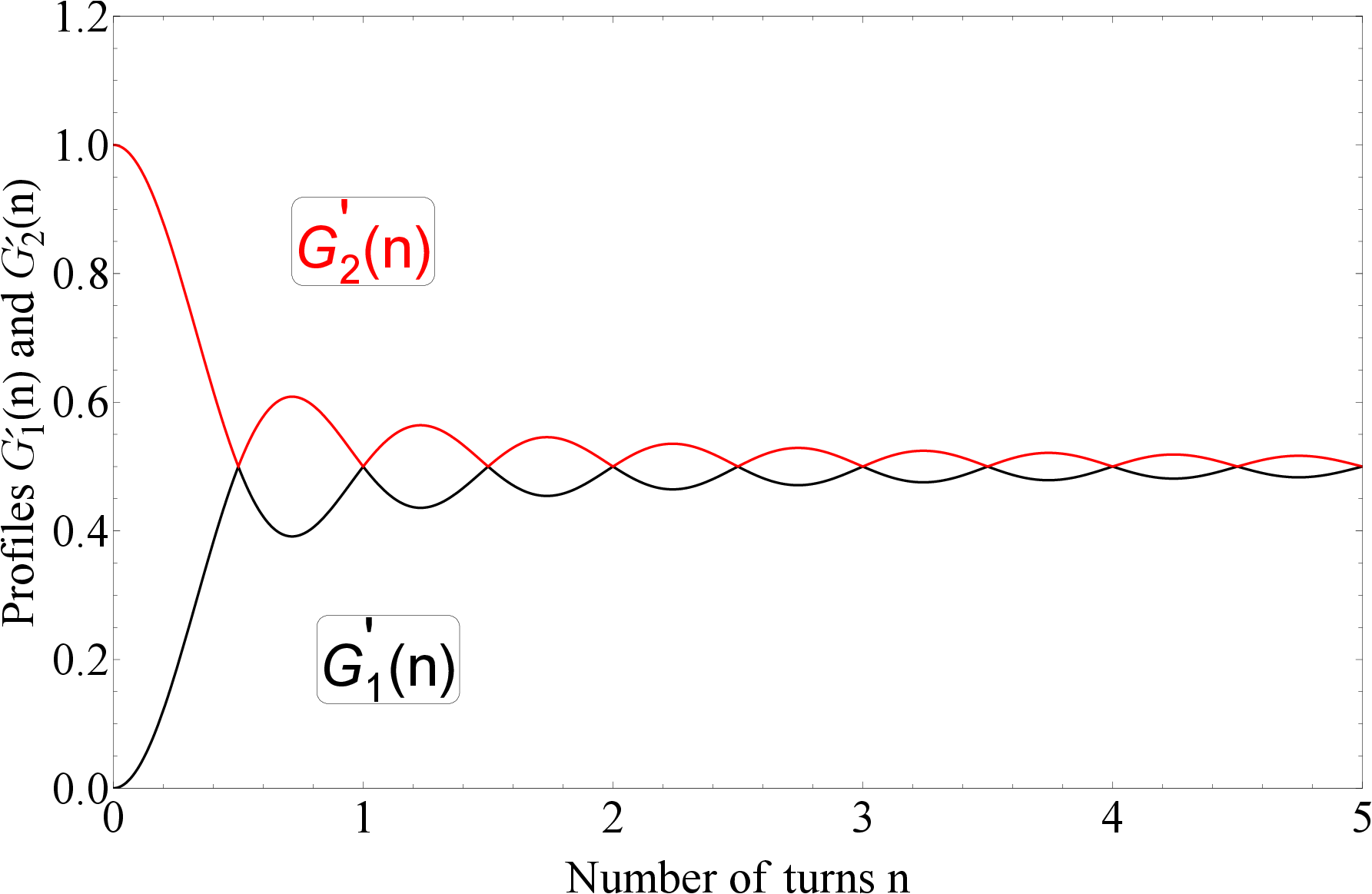}} 
\et
\caption{
$G'_1(n)$ and $G'_2 (n)$ profile dependences on the number of turns $n$ (the big ellipse on resonance).
}
\label{fig:G1G2nshtrikh}
\end{figure}
%***************************************************************

%***************************************************************
\begin{figure}[t]
\bt{r}
\centerline{\includegraphics[width=1\columnwidth]{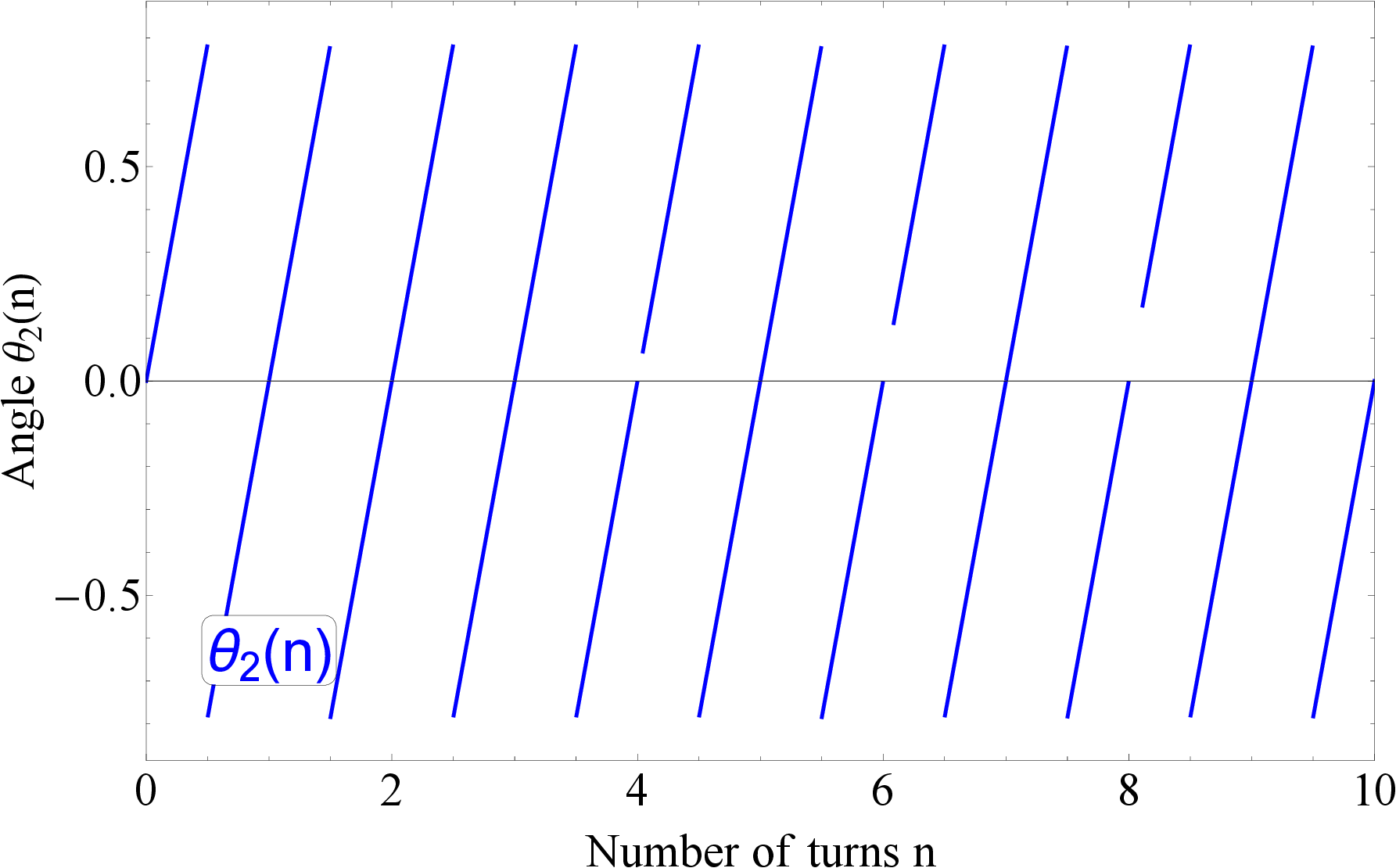}} 
\et
\caption{
Angle $\theta_2(n)$ dependence on the number of turns $n$ (the large ellipse on resonance).
}
\label{fig:theta2n}
\end{figure}
%***************************************************************

In the limit $\omega \rightarrow \omega_c$, the angle of the first ellipse (see Fig.~\ref{fig:theta1n}) is equal to
\be
\theta_1 = \frac{1}{2} \arctan{\left(\frac{a_c - \sin{a_c}}{\cos{a_c} - 1}\right)},
\ee
which has its extrema at $n \rightarrow 0$
\be
\theta_{1,\text{max}} = \lim\limits_{n \to 0} \theta_1 (n) = 0, 
\ee
and at $n \rightarrow k + 1/2, \quad k \in \mathbb{N}_0$ and $n \rightarrow \infty$
\be
\theta_{1,\text{min}} = \lim\limits_{n \to k + 1/2} \theta_1 (n) = \lim\limits_{n \to \infty} \theta_1 (n) = - \frac{\pi}{2}. 
\ee

%%%%%%%%%%%%%%%%%%%%%%%%%%%%%%%%%%%%%%%%%%%%%%%%%%%%%%%%%%%%%%%%%%%%%%%%%%%%%%%%%%%%%%%%%%%%%%%%%%%%%%%%%%%%%%%%%%%%%%%%%%%%%%%%%%%%%%%%%
%%%%%%%%%%%%%%%%%%%%%%%%%%%%%%%%%%%%%%%%%%%%%%%%%%%%%%%%%%%%%%%%%%%%%%%%%%%%%%%%%%%%%%%%%%%%%%%%%%%%%%%%%%%%%%%%%%%%%%%%%%%%%%%%%%%%%%%%%
\subsection{TOF between a deflector and a detector screen: on resonance\label{Subsec:TOFdefscr}}
%%%%%%%%%%%%%%%%%%%%%%%%%%%%%%%%%%%%%%%%%%%%%%%%%%%%%%%%%%%%%%%%%%%%%%%%%%%%%%%%%%%%%%%%%%%%%%%%%%%%%%%%%%%%%%%%%%%%%%%%%%%%%%%%%%%%%%%%%
%%%%%%%%%%%%%%%%%%%%%%%%%%%%%%%%%%%%%%%%%%%%%%%%%%%%%%%%%%%%%%%%%%%%%%%%%%%%%%%%%%%%%%%%%%%%%%%%%%%%%%%%%%%%%%%%%%%%%%%%%%%%%%%%%%%%%%%%%

In the limit $\omega \rightarrow \omega_c \Rightarrow \{X^\prime \rightarrow X^\prime_c, Y^\prime \rightarrow Y^\prime_c\}$ 
\be
\begin{gathered}
X^\prime_c = \tau_D \left(A^\prime_{1c} c_\phi + B^\prime_{1c} s_\phi \right) = a^\prime_{1c} c_\phi + b^\prime_{1c} s_\phi = \\ 
= A^\prime_{xc} \cos{\left(\phi + \phi^\prime_{xc}\right)}, \\
Y^\prime_c = \tau_D \left(A^\prime_{2c} c_\phi + B^\prime_{2c} s_\phi\right) = a^\prime_{2c} c_\phi + b^\prime_{2c} s_\phi = \\
= A^\prime_{yc} \sin{\left(\phi + \phi^\prime_{yc}\right)}, \\ 
a^\prime_{1c} = b^\prime_{2c} = \frac{A(\tilde{E}_\perp)}{8 \omega^2_c} \left(1 - \cos{a_c}\right) a_D, \\
b^\prime_{1c} = \frac{A(\tilde{E}_\perp)}{8 \omega^2_c} \left(a_c + \sin{a_c}\right) a_D,\\
a^\prime_{2c} = \frac{A(\tilde{E}_\perp)}{8 \omega^2_c} \left(a_c - \sin{a_c}\right) a_D.
\end{gathered}
\ee

\begin{widetext}
The sizes of the big ellipse are the following
\bse\label{Shamaevellipse}
\begin{gather}
\chi^\prime_{c} = \frac{\abs{A(\tilde{E}_\perp)}}{8 \omega^2_c} a_D a_c \left(1 - \abs{\frac{\sin{(a_c/2)}}{a_c /2}} \right) = \frac{\abs{A(\tilde{E}_\perp)} l L_D}{v_z^2} G'_1 (n), \label{Shamaevform1}\\
\varkappa^\prime_{c} = \frac{\abs{A(\tilde{E}_\perp)}}{8 \omega^2_c} a_D a_c \left(1 + \abs{\frac{\sin{(a_c/2)}}{a_c /2}} \right) = \frac{\abs{A(\tilde{E}_\perp)} l L_D}{v_z^2} G'_2 (n), \label{Shamaevform2} \\
G'_1 (n) = \frac{1}{2} \left(1 - \abs{\frac{\sin{2\pi n}}{2 \pi n}} \right), \quad G'_2 (n) = \frac{1}{2} \left(1 + \abs{\frac{\sin{2\pi n}}{2 \pi n}} \right),
\end{gather}
\ese 
where the final formulas ($\sinc{(x)} = \sin{x}/x$ function) in \eqref{Shamaevform1} and \eqref{Shamaevform2} exactly correspond to the ones in the textbook \cite{ZhigarevBook} (note that in the textbook $n=1$). 
\end{widetext}

In the limit $\omega \rightarrow \omega_c$, the angle of the big ellipse (see Fig.~\ref{fig:theta2n}) is equal to
\be
\begin{gathered}
\theta_2 = \frac{1}{2} \arctan{\left(\tan{\frac{a_c}{2}}\right)} = \frac{a_c}{4} - \frac{\pi}{2} \, \left\lceil \frac{1}{2}\,\left\lfloor \frac{a_c}{\pi} \right\rfloor \right\rceil = \\
= \pi n - \frac{\pi}{2} \, \left\lceil \frac{1}{2}\,\left\lfloor 4 n \right\rfloor \right\rceil,
\end{gathered}
\ee
where, to restrict $\theta_2$ to the first quadrant, the ceiling $\lceil * \rceil$ and the floor $\lfloor * \rfloor$ rounding functions are used. This is a mathematical convention and in physics it is usually more natural to use rotation angle which is continuous function of $n$, i.e., $\theta_2 = \pi n$.

Hereby, the Shamaev conditions to get a circle simply can be seen: when the following
\begin{enumerate}
\item Shamaev resonance $\omega \rightarrow \omega_c$, i.e., the equality of the TOF ($\tau_c = \tfrac{\lambda}{v_z} = \tfrac{2\pi}{\omega_c}$) of a charged particle through one turn of the helix to the period ($T = \tfrac{2\pi}{\omega}$) of the high-frequency voltage: $\tau_c = T$, 
\item a half-integer number of turns $n = k/2 \quad (a_c / 2  = 2\pi k), \quad k \in \mathbb{N}$, which is equivalent to $\tau = \tfrac{\pi k}{\omega_c}$ written in the textbook, \cite{ZhigarevBook}
\end{enumerate} 
hold, the big ellipse turns to a circle ($\chi^\prime_{c} = \varkappa^\prime_{c}$) with a radius (see Fig.~\ref{fig:G1G2nshtrikh})
\be\label{radiusrprimec}
r^\prime_c = \frac{\abs{A(\tilde{E}_\perp)}}{8 \omega^2_c} a_c a_D = \frac{e \tilde{E}_\perp}{m\gamma} \frac{2 \pi^2 n n_D}{\omega^2_c} = \frac{e \tilde{E}_\perp}{m\gamma} \frac{l L_D}{2 v^2_z}.
\ee

%%%%%%%%%%%%%%%%%%%%%%%%%%%%%%%%%%%%%%%%%%%%%%%%%%%%%%%%%%%%%%%%%%%%%%%%%%%%%%%%%%%%%%%%%%%%%%%%%%%%%%%%%%%%%%%%%%%%%%%%%%%%%%%%%%%%%%%%%
%%%%%%%%%%%%%%%%%%%%%%%%%%%%%%%%%%%%%%%%%%%%%%%%%%%%%%%%%%%%%%%%%%%%%%%%%%%%%%%%%%%%%%%%%%%%%%%%%%%%%%%%%%%%%%%%%%%%%%%%%%%%%%%%%%%%%%%%%
\subsection{Total TOF: on resonance\label{Subsec:TOFdefscr}}
%%%%%%%%%%%%%%%%%%%%%%%%%%%%%%%%%%%%%%%%%%%%%%%%%%%%%%%%%%%%%%%%%%%%%%%%%%%%%%%%%%%%%%%%%%%%%%%%%%%%%%%%%%%%%%%%%%%%%%%%%%%%%%%%%%%%%%%%%
%%%%%%%%%%%%%%%%%%%%%%%%%%%%%%%%%%%%%%%%%%%%%%%%%%%%%%%%%%%%%%%%%%%%%%%%%%%%%%%%%%%%%%%%%%%%%%%%%%%%%%%%%%%%%%%%%%%%%%%%%%%%%%%%%%%%%%%%%

In the limit $\omega \rightarrow \omega_c \Rightarrow \{X \rightarrow X_c, Y \rightarrow Y_c\}$ 
\be
\begin{gathered}
X_c = A_{1c} c_\phi + B_{1c} s_\phi, \\
Y_c = A_{2c} c_\phi + B_{2c} s_\phi, \\ 
A_{1c} = B_{2c} = \frac{A(\tilde{E}_\perp)}{8 \omega^2_c} \left(a_c - \sin{a_c} + a_D (1 - \cos{a_c}) \right), \\
B_{1c} = \frac{A(\tilde{E}_\perp)}{8 \omega^2_c} \left(\frac{a^2_c}{2} + 1 - \cos{a_c} + a_D (a_c + \sin{a_c}) \right), \\
A_{2c} = \frac{A(\tilde{E}_\perp)}{8 \omega^2_c} \left(\frac{a^2_c}{2} - (1 - \cos{a_c}) + a_D (a_c - \sin{a_c}) \right).
\end{gathered}
\ee

\begin{widetext}
The sizes of the resulting ellipse are the following
\bse\label{finalellipse}
\begin{gather}
\chi_{c} = \frac{\abs{A(\tilde{E}_\perp)}}{16 \omega^2_c} \frac{\abs{f_1 (a_c, a_D) - k(a_c)}}{\sqrt{k(a_c) + f_2 (a_c, a_D) + l(a_c, a_D)}} = \frac{e \tilde{E}_\perp}{m \gamma} \frac{l^2}{4 v^2_z} \frac{K_3 (n) \abs{K_1 (n, n_D) - \frac{1}{(4 \pi n)^2}}}{\sqrt{K_{2+} (n, n_D) + \frac{1}{(4\pi n)^2}}} = \frac{l L_D \abs{A(\tilde{E}_\perp)}}{v^2_z} G_1 (n, n_D), \\
\varkappa_{c} = \frac{\abs{A(\tilde{E}_\perp)}}{16 \omega^2_c} \frac{\abs{f_1 (a_c, a_D) - k(a_c)}}{\sqrt{k(a_c) + f_2 (a_c, a_D) - l(a_c, a_D)}} = \frac{e \tilde{E}_\perp}{m \gamma} \frac{l^2}{4 v^2_z} \frac{K_3 (n) \abs{K_1 (n, n_D) - \frac{1}{(4 \pi n)^2}}}{\sqrt{K_{2-} (n, n_D) + \frac{1}{(4\pi n)^2}}} = \frac{l L_D \abs{A(\tilde{E}_\perp)}}{v^2_z} G_2 (n, n_D), \\
l(a_c, a_D) = 4 a_c (a_c + 2 a_D) \sqrt{2 a_D (a_c + a_D) (1 - \cos{a_c}) + 2 (1 - \cos{a_c}) - 2 a_c \sin{a_c} + a_c^2}, \\
f_1 (a_c, a_D) = a_c^4 + 4 a_D (a_c + a_D) (2 (\cos{a_c} - 1) + a_c^2), \\
f_2 (a_c, a_D) = a_c^4 + 4 a_D (a_c + a_D) (2 (1 - \cos{a_c}) + a_c^2), \\
k (a_c) = 8 (1 - \cos{a_c}) - 4 a_c (2 \sin{a_c} - a_c), \\ 
K_1 (n, n_D) = \frac{1 + 4 \frac{n_D}{n} \left(1 + \frac{n_D}{n} \right) \left(2 \frac{\cos{4 \pi n} - 1}{(4 \pi n)^2} + 1 \right) }{8 \frac{1-\cos{4 \pi n}}{(4\pi n)^2} - 4 \left(2 \frac{\sin{4 \pi n}}{4 \pi n} - 1 \right)}, \\
K_{2\pm}(n,n_D) = \frac{1 + 4\frac{n_D}{n}\Big(1+\frac{n_D}{n}\Big)\Big(1 + 2\frac{1-\cos(4\pi n)}{(4\pi n)^2}\Big) \pm \tilde l(n,n_D)}{8\frac{1-\cos(4\pi n)}{(4\pi n)^2} - 4\Big(2\frac{\sin(4\pi n)}{4\pi n}-1\Big)}, \\
K_3 (n) = \sqrt{8 \frac{1 - \cos{4\pi n}}{(4 \pi n)^2} - 4 \left(2 \frac{\sin{4 \pi n}}{4 \pi n} - 1 \right)}, \\
\tilde{l} (n, n_D ) = 4 \left(1 + 2 \frac{n_D}{n}\right) \sqrt{2 \frac{n_D}{n} \left(1 + \frac{n_D}{n} \right) \frac{1 - \cos{4 \pi n}}{(4 \pi n)^2} + 2 \frac{1 - \cos{4 \pi n}}{(4\pi n)^4} - 2 \frac{\sin{4\pi n}}{(4\pi n)^3} + \frac{1}{(4\pi n)^2}}, \\
G_1 (n, n_D) = \frac{1}{4} \frac{n}{n_D} \frac{K_3 (n) \abs{K_1 (n, n_D) - \frac{1}{(4 \pi n)^2}}}{\sqrt{K_{2+} (n, n_D) + \frac{1}{(4\pi n)^2}}}, \\
G_2 (n, n_D) = \frac{1}{4} \frac{n}{n_D} \frac{K_3 (n) \abs{K_1 (n, n_D) - \frac{1}{(4 \pi n)^2}}}{\sqrt{K_{2-} (n, n_D) + \frac{1}{(4\pi n)^2}}},
\end{gather}
\ese 
\end{widetext}

\begin{figure*}[t]
  \centering
  
  \begin{subfigure}[t]{0.48\textwidth}
    \centering
    \includegraphics[width=\linewidth]{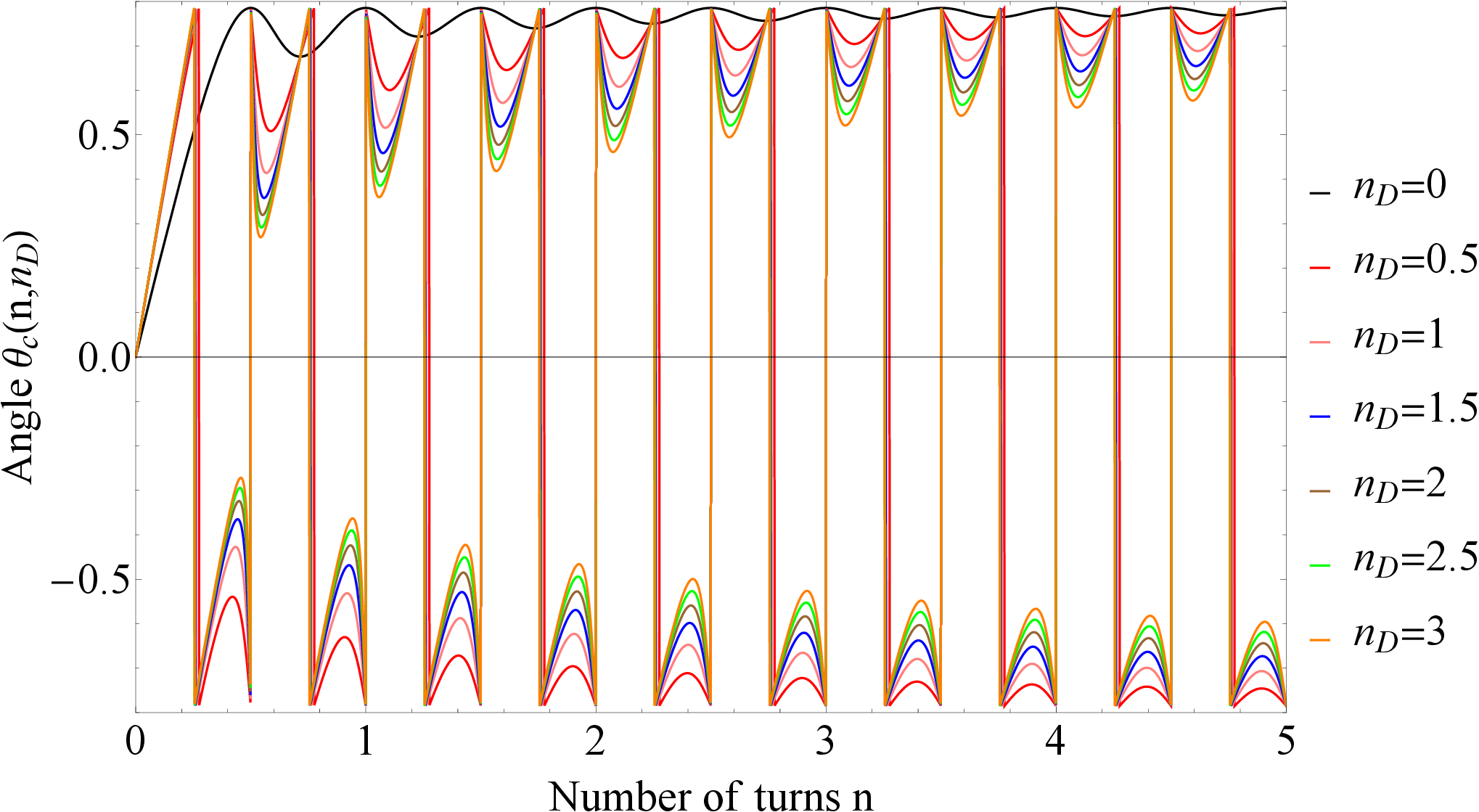}
  \end{subfigure}
  \vfill
  \begin{subfigure}[t]{0.48\textwidth}
    \centering
    \includegraphics[width=\linewidth]{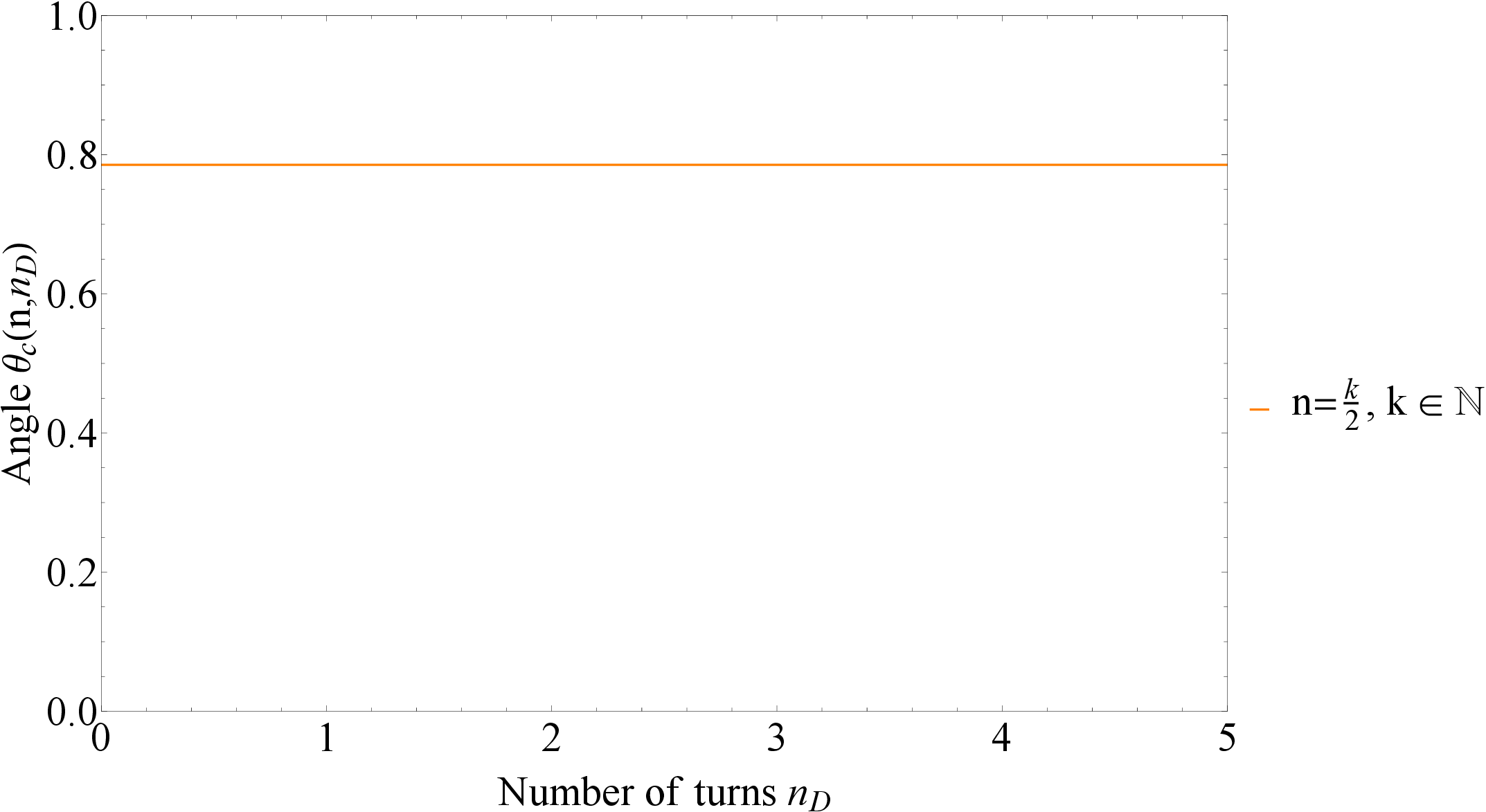}
  \end{subfigure}
  
  \caption{${\theta}_c(n,n_D)$ angle dependences (a) on the number of turns $n$ for the fixed values of the \emph{phantom} number of turns $n_D$, and (b) on the \emph{phantom} number of turns $n_D$ for the fixed values of the number of turns $n$ (the resulting ellipse on resonance).}
  \label{fig:thetacnnD}
\end{figure*}

In the limit $\omega \rightarrow \omega_c$, the angle of the final (the resulting) ellipse (see Fig.~\ref{fig:thetacnnD}) is equal to
\be
\begin{gathered}
\theta_c = \frac12 \arctan{\left(\frac{a_c - \sin{a_c} + a_D (1 - \cos{a_c})}{1 - \cos{a_c} + a_D \sin{a_c}}\right)}.
\end{gathered}
\ee

\begin{figure*}[t]
  \centering
  
  \begin{subfigure}[t]{0.48\textwidth}
    \centering
    \includegraphics[width=\linewidth]{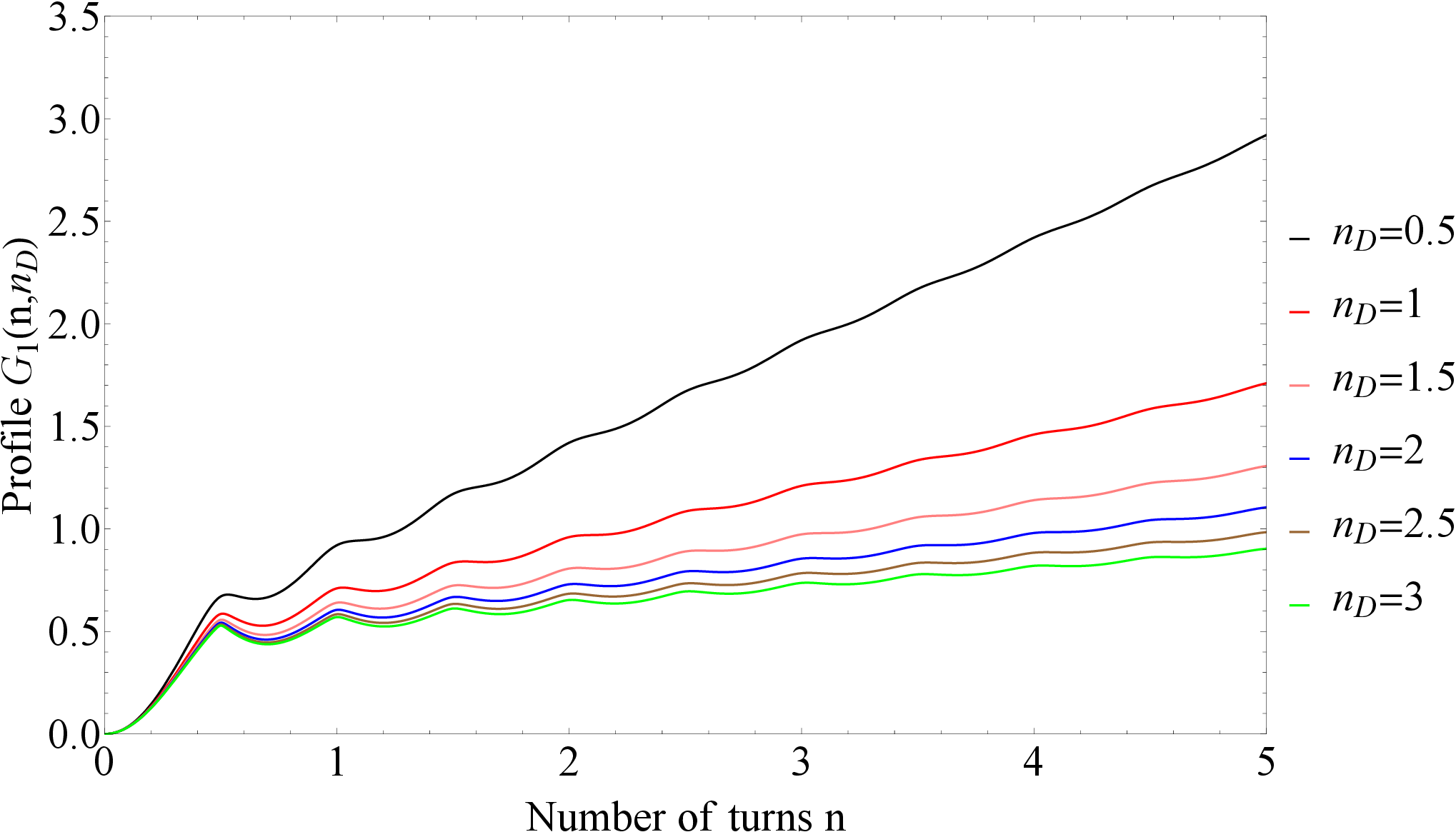}
    \subcaption{${G}_1(n,n_D)$ profile.}
    \label{fig:G1nnDvsn}
  \end{subfigure}
  \vfill
  \begin{subfigure}[t]{0.48\textwidth}
    \centering
    \includegraphics[width=\linewidth]{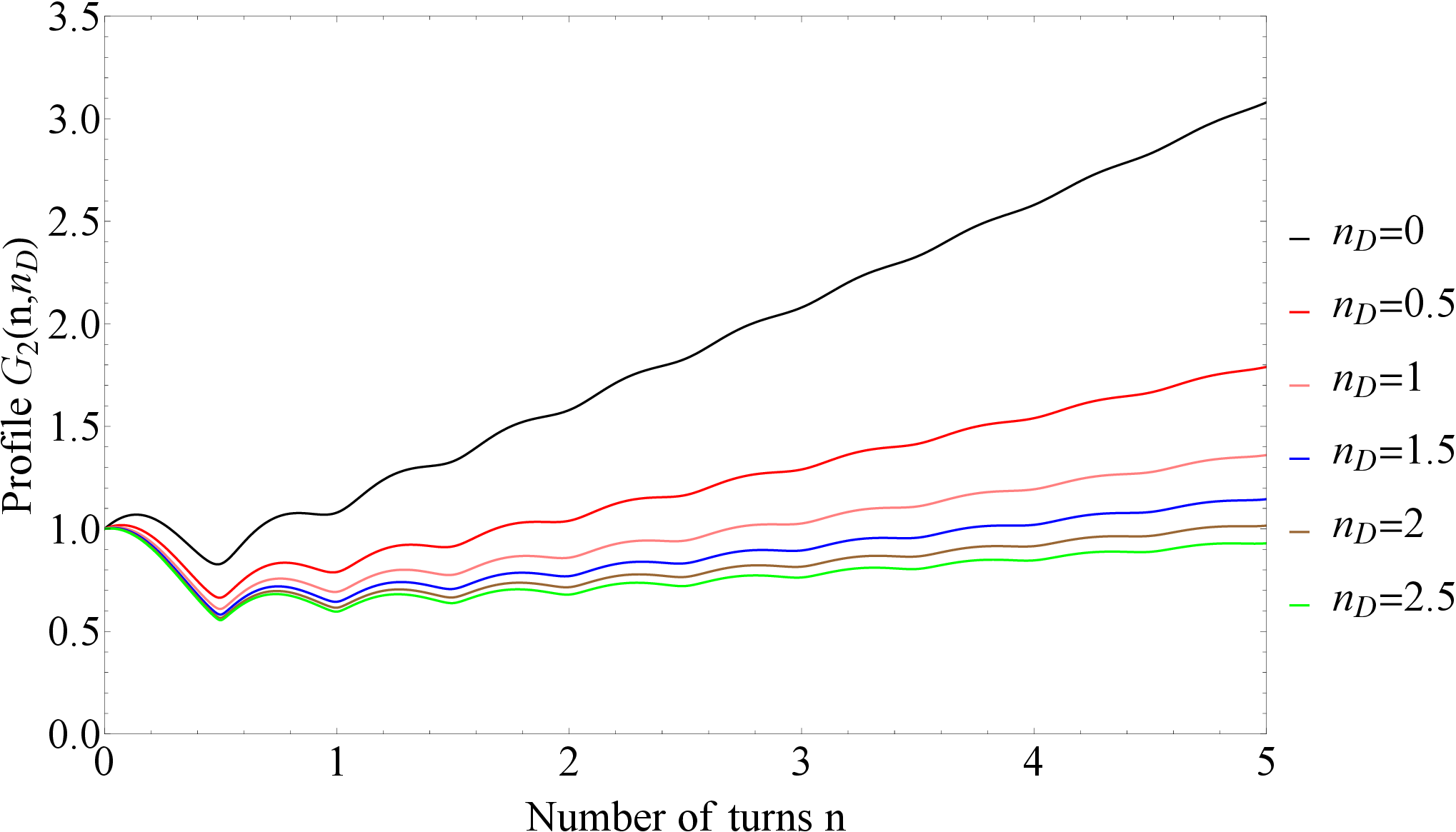}
    \subcaption{${G}_2(n,n_D)$ profile.}
    \label{fig:G2nnDvsn}
  \end{subfigure}
  
  \caption{${G}_1(n,n_D)$ and ${G}_2 (n,n_D)$ profile dependences on the number of turns $n$ for the fixed values of the number of \emph{phantom} turns $n_D$ (the resulting ellipse on resonance).}
  \label{FIG:G1G2nnDforfixednD}
\end{figure*}

\begin{figure*}[t]
  \centering
  
  \begin{subfigure}[t]{0.48\textwidth}
    \centering
    \includegraphics[width=\linewidth]{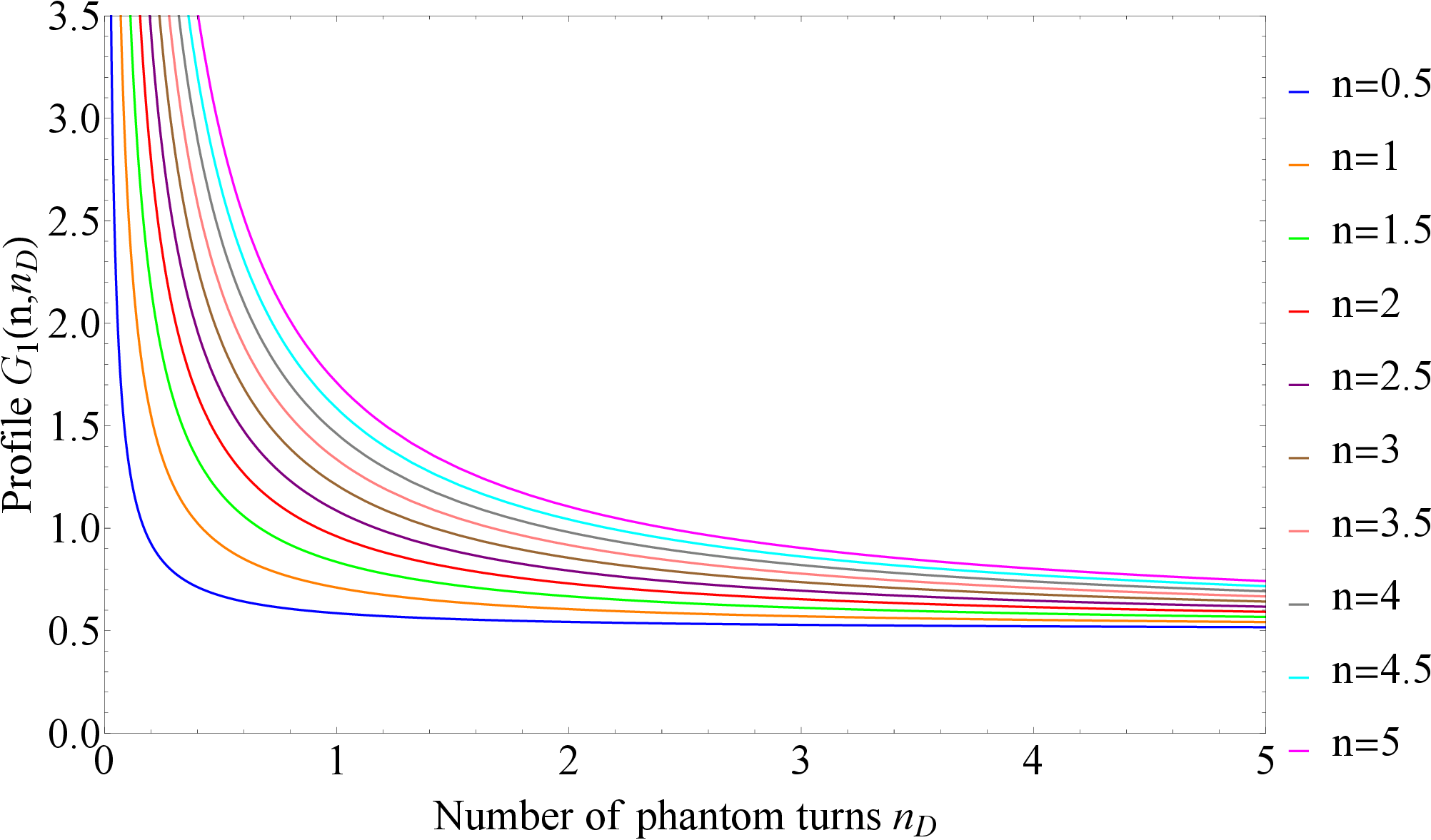}
    \subcaption{${G}_1(n,n_D)$ profile.}
    \label{fig:G1nnDvsn}
  \end{subfigure}
  \vfill
  \begin{subfigure}[t]{0.48\textwidth}
    \centering
    \includegraphics[width=\linewidth]{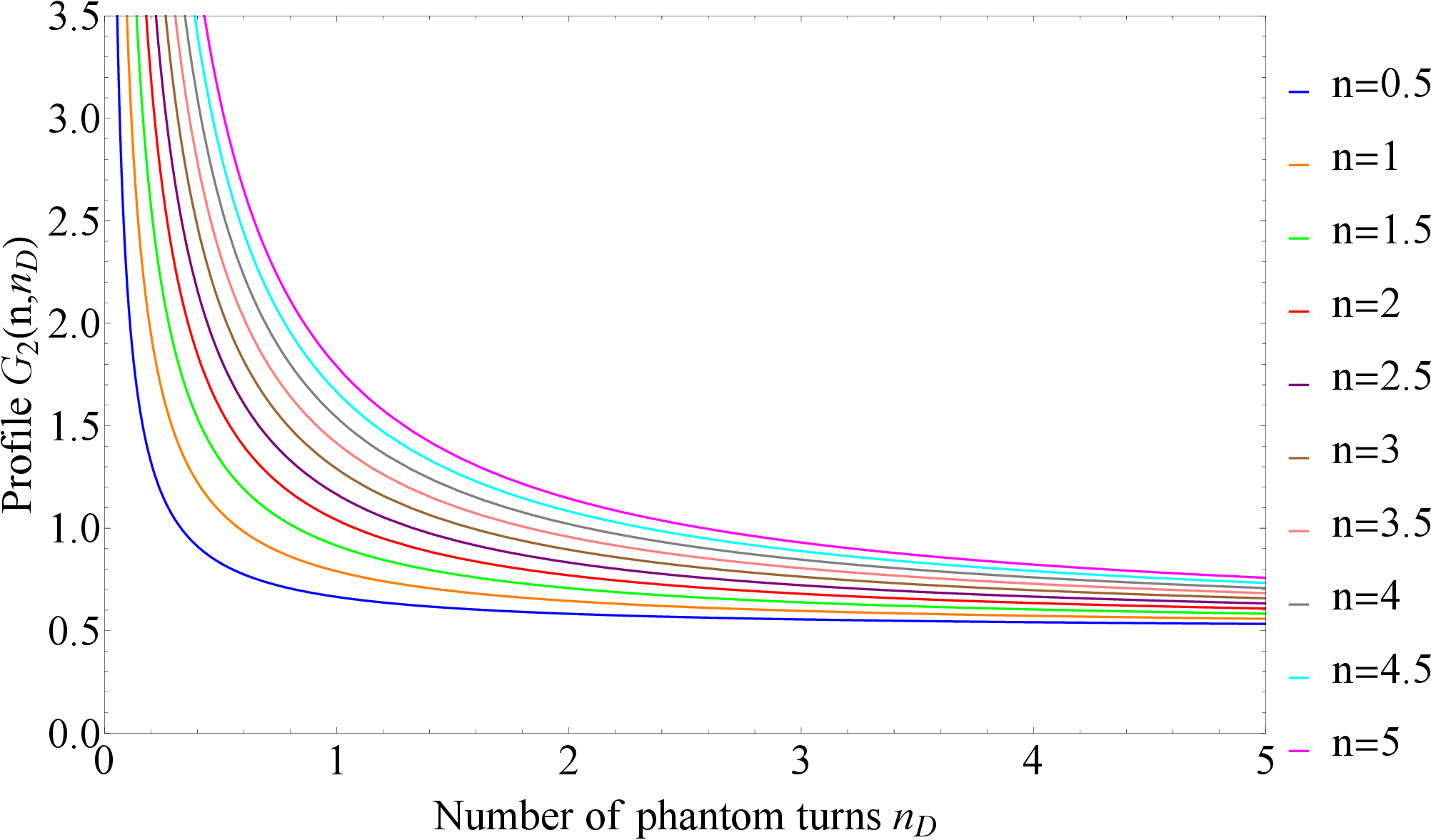}
    \subcaption{${G}_2(n,n_D)$ profile.}
    \label{fig:G2nnDvsn}
  \end{subfigure}
  
  \caption{${G}_1(n,n_D)$ and ${G}_2 (n,n_D)$ profile dependences on the \emph{phantom} number of turns $n_D$ for the fixed values of the number of turns $n$ (the resulting ellipse on resonance).}
  \label{FIG:G1G2nnDforfixedn}
\end{figure*}

Figs.~\ref{FIG:G1G2nnDforfixednD} and~\ref{FIG:G1G2nnDforfixedn} show the resulting ellipse semi-axes' (${G}_1(n,n_D)$ and ${G}_2 (n,n_D)$) profile dependences on the number of turns $n$ for the fixed values of the number of \emph{phantom} turns $n_D$ (see Fig.~\ref{FIG:G1G2nnDforfixednD}) and on the \emph{phantom} number of turns $n_D$ for the fixed values of the number of turns $n$ (see Fig.~\ref{FIG:G1G2nnDforfixedn}), respectively. In the far limit $n_D \rightarrow \infty$ from the deflector, the ellipse sizes tend to 

\be
\begin{gathered}
\lim\limits_{n_D \to \infty} G_1 (n, n_D) = \frac{1}{2} 
\left(1 - \frac{1}{\sqrt{2}} \frac{\sin^2{2\pi n}}{(2\pi n)^2}\right), \\
\lim\limits_{n_D \to \infty} G_2 (n, n_D) = \frac{1}{2} \left(1 + \frac{1}{\sqrt{2}} \frac{\sin^2{2\pi n}}{(2\pi n)^2}\right).
\end{gathered}
\ee
This asymptotic behaviour can be seen in Fig.~\ref{FIG:G1G2nnDforfixedn}. So, for sufficiently large distance between deflector and detector screen $n_D \gg 1$ and in the case of Shamaev conditions $\omega \rightarrow \omega_c$ and $n = k/2, \quad k \in \mathbb{N}$, ellipse tends to a circle with a radius $r_c = \frac{l L_D \abs{A(\tilde{E}_\perp)}}{2 v^2_z}$ ($\lim\limits_{n_D \to \infty} G_1 (k/2, n_D) = \lim\limits_{n_D \to \infty} G_2 (k/2, n_D) = 1/2$) equal to $r^\prime_c$ (see Eq.~\eqref{radiusrprimec}).

 %Top resultsComparison Shamaev and Theory.nb on Desktop (even in overleaf file)

 %Top resultsComparison Shamaev and Theory.nb on Desktop (even in overleaf file)

 %Top resultsComparison Shamaev and Theory.nb on Desktop (even in overleaf file)

% \\
% angle \\
% $r^\prime_c$ \\
% graphics for r^\prime_c as in Fig. 3\\

% To get a circle $x_c$ and $y_c$ must have equal amplitudes $a_{yc} - a_{xc} = 0$ (see Eq.~\eqref{eq.1}) and their cosine and sine functions must be in phase $\phi_{yc} - \phi_{xc} = \pi k, \quad k \in \mathbb{Z}$ (see Eq.~\eqref{eq.2})
% \bse
% \begin{gather}
% \left(a^2_{1c} + b^2_{1c}\right) - \left(a^2_{2c} + a^2_{1c}\right) = 0, \label{eq.1} \\
% a^2_{1c} - a_{2c} b_{1c} = 0. \label{eq.2}
% \end{gather}
% \ese

% Eq.~\eqref{eq.1} holds when $a_{2c} = b_{1c}$ $\Rightarrow$ $a_c = \pi k, \quad k \in \mathbb{Z}$ and Eq.~\eqref{eq.2}: when the function $g(a_c) = - \tfrac{a^4_c}{4} + a^2_c - 2 a_c \sin{a_c} + 2 (1 - \cos{a_c}) = 0$ $\Rightarrow$ $a_c \rightarrow 0$.

% In this way, as much as the number of turns is small, the ellipse will get closer to the circle, but never this limit $a_c \rightarrow 0$ of a circle exists, because naturally the deflector must have some number of turns.

% Non-preferred case can be the limit of a line. This happens when cosine
% and sine functions have $\pi/2$ phase shift between them:
% $\phi_{yc} - \phi_{xc} = \pi/2 + \pi k, \quad k \in \mathbb{Z}$, which leads to $b_{1c} a_{yc} \pm b_{2c} a_{xc} = 0$.

%%%%%%%%%%%%%%%%%%%%%%%%%%%%%%%%%%%%%%%%%%%%%%%%%%%%%%%%%%%%%%%%%%%%%%%%%%%%%%%%%%%%%%%%%%%%%%%%%%%%%%%%%%%%%%%%%%%%%%%%%%%%%%%%%%%%%%%%%
%%%%%%%%%%%%%%%%%%%%%%%%%%%%%%%%%%%%%%%%%%%%%%%%%%%%%%%%%%%%%%%%%%%%%%%%%%%%%%%%%%%%%%%%%%%%%%%%%%%%%%%%%%%%%%%%%%%%%%%%%%%%%%%%%%%%%%%%%
\section{To get closer to the circle\label{Sec:getcircle}}
%%%%%%%%%%%%%%%%%%%%%%%%%%%%%%%%%%%%%%%%%%%%%%%%%%%%%%%%%%%%%%%%%%%%%%%%%%%%%%%%%%%%%%%%%%%%%%%%%%%%%%%%%%%%%%%%%%%%%%%%%%%%%%%%%%%%%%%%%
%%%%%%%%%%%%%%%%%%%%%%%%%%%%%%%%%%%%%%%%%%%%%%%%%%%%%%%%%%%%%%%%%%%%%%%%%%%%%%%%%%%%%%%%%%%%%%%%%%%%%%%%%%%%%%%%%%%%%%%%%%%%%%%%%%%%%%%%%

Although it is written in the textbook that there is a way to get a circle on the screen, I argue that claim. In the textbook, it is assumed that electrons are deflected from the deflector's axis at the end of the helical deflector (a full \emph{paraxial approximation}). In reality, if one imagines that the screen is exactly after the deflector, the electrons will be incident to the screen painting a small ellipse, which opposes to the opinion in the textbook. This small ellipse; since at the end of deflector the transverse radius and the transverse velocity have different directions, i.e., $\tan{\alpha} \neq \tan{\beta}$ in general ($\tan{\alpha} = y(\tau)/x(\tau)$, $\tan{\beta} = v_y(\tau)/v_x(\tau)$); can drastically change the final ellipse (TOF between screen and the start of deflector), which will be a merging of a small ellipse (TOF at deflector) and a Shamaev ellipse (TOF between screen and the end of deflector). Although the Shamaev ellipse becomes a circle in the case of Shamaev condition (Shamaev resonance $\omega \rightarrow \omega_c$ and $\tau = \pi k / \omega_c$ or $n = k/2, \quad k \in \mathbb{N}$), the small ellipse is not negligible (compare Forms.~\eqref{Shamaevellipse} and~\eqref{finalellipse}). Moreover, if one makes comparison between \emph{capacitor model} and \emph{textbook model}, in the case of Shamaev conditions and a \emph{capacitor model}, the ellipse gets closer to the circle, but in any case it remains as an ellipse (see Subfigs.~\ref{fig:Capacitor_vs_textbook_k=1_n=1} and~\ref{fig:Capacitor_vs_textbook_k=1_n=0.5}). The reason is a non-zero electric field before and after the deflector (close to the edges).

% Anyway, the coordinates can be represented as
% \be
% \begin{gathered}
% X = A_1 c_\phi + B_1 s_\phi = A_x c_{\phi+\phi_X}, \\
% Y = A_2 c_\phi + B_2 s_\phi = A_y c_{\phi+\phi_Y}, \\
% \end{gathered}
% \ee
% and the ellipse will get closer to a circle, by minimizing the following loss function of optimization
% \bse\label{method-1}
% \begin{gather}
% \mathcal{E} = \lambda_A \mathcal{E}_A + \lambda_\phi \mathcal{E}_\phi \rightarrow \text{min}, \\
% \mathcal{E}_A = \abs{\left(A^2_1 + B^2_1\right) - \left(A^2_2 + B^2_2\right)}^2, \\
% \mathcal{E}_\phi = \abs{A_1 B_2 - A2 B1}^2,
% \end{gather}
% \ese
% where $\mathcal{E}_A$ is responsible for minimization of the difference in the amplitudes $\Delta A = A_x - A_y \rightarrow \text{min}$ and $\mathcal{E}_A$ for the difference in the phases $\Delta \phi = \phi_X - \phi_Y \rightarrow \text{min}$. A common procedure will be to take regularization weights equal $\lambda_A = \lambda_\phi = 1$, but for generality they are given as arbitraries. For fast computation of the results the \emph{least mean square} form, i.e, the sum of the squares of the absolute errors is taken.  

%%%%%%%%%%%%%%%%%%%%%%%%%%%%%%%%%%%%%%%%%%%%%%%%%%%%%%%%%%%%%%%%%%%%%%%%%%%%%%%%%%%%%%%%%%%%%%%%%%%%%%%%%%%%%%%%%%%%%%%%%%%%%%%%%%%%%%%%%
%%%%%%%%%%%%%%%%%%%%%%%%%%%%%%%%%%%%%%%%%%%%%%%%%%%%%%%%%%%%%%%%%%%%%%%%%%%%%%%%%%%%%%%%%%%%%%%%%%%%%%%%%%%%%%%%%%%%%%%%%%%%%%%%%%%%%%%%%
\section{\emph{On and off} single-electron pencil beam\label{Sec:pencilbeam}}
%%%%%%%%%%%%%%%%%%%%%%%%%%%%%%%%%%%%%%%%%%%%%%%%%%%%%%%%%%%%%%%%%%%%%%%%%%%%%%%%%%%%%%%%%%%%%%%%%%%%%%%%%%%%%%%%%%%%%%%%%%%%%%%%%%%%%%%%%
%%%%%%%%%%%%%%%%%%%%%%%%%%%%%%%%%%%%%%%%%%%%%%%%%%%%%%%%%%%%%%%%%%%%%%%%%%%%%%%%%%%%%%%%%%%%%%%%%%%%%%%%%%%%%%%%%%%%%%%%%%%%%%%%%%%%%%%%%

It will be difficult or ineffective to handle all the mentioned conditions. Therefore, it is important to create a framework to work with the ellipse. The initial equations provide that framework. 

Let's imagine that we have \emph{on and off} single-electron pencil beam, which is an approximation of the bunched beam. For this beam of charged particles, their track on the ellipse will be partial ellipse arcs with unequal lengths, when in the case of a circle limit the partial arc lengths should be equal. The question appears --- \emph{How to measure TOF using these unequal lengths?} Experimental answer will be to use RMS (root-mean-square) measure. 

Let's imagine that \emph{on and off} single-electron pencil beam enters the deflector with \emph{on}-beam time intervals: 

\begin{equation}
\{t_1,t_2\}, \{t_3,t_4\}, \ldots, \{t_{N-1},t_N\}  
\end{equation}
(with the time lengths $T_{on} = t_2 - t_1 = t_4 - t_3 = \cdots = 
t_N - t_{N-1}$), and \emph{off}-beam time intervals: 
\begin{equation}
\{t_2,t_3\}, \{t_4,t_5\}, \ldots, \{t_{N},t_{N+1}\}  
\end{equation}
(with the time lengths $T_{off} = t_3 - t_2 = t_5 - t_4 = \cdots = t_{N+1} - t_N$), where $N$ is an even number and the corresponding phases
are $\phi_i = \phi (t_{i})$. Let's imagine also that $\phi_{N k +1} = \phi_1 + 2\pi k, \quad k \in \mathbb{N}$ and, hence, due to this periodicity the pattern on the screen is repetitive and the partial ellipse arcs coincide after one period $T = t_{N+1}-t_1 = N (T_{on} + T_{off}) / 2 = 2\pi/\omega$ (without overlap or gaps). This is crucial for a time measurement of \emph{on}-beam lengths and for the work of an apparatus as a timing device (timer).

\emph{On}-beam lengths can be calculated according to the Eq.~\eqref{lengthinellipse} as $s_i = \int\limits_{\phi_i}^{\phi_{i+1}} \sqrt{\left(\frac{dX}{d\phi}\right)^2 + \left(\frac{dY}{d\phi}\right)^2} d\phi$. RMS of \emph{on}-beam lengths will give length characteristic of time of \emph{on}-beam
\begin{equation}
s_{on} = \sqrt{\frac{2}{N}\sum\limits_{i = 1, odd}^{N-1} s^2_i}. 
\end{equation}
Correspondingly, for \emph{off}-beam:
\begin{equation}
s_{off} = \sqrt{\frac{2}{N}\sum\limits_{i = 2, even}^{N} s^2_i}.
\end{equation}
Another characteristic length of time can be simply average. For \emph{on}-beam:
\be
l_{on} = \frac{2}{N}\sum\limits_{i = 1, odd}^{N-1} s_i,
\ee
and for \emph{off}-beam:
\begin{equation}
l_{off} = \frac{2}{N}\sum\limits_{i = 2, even}^{N} s_i. 
\end{equation}

Finally, their sum $\tfrac{N}{2}(l_{on} + l_{off})$ will give the length of ellipse (perimeter)
\begin{equation}
\begin{gathered}
\frac{N}{2}(l_{on} + l_{off}) = \sum\limits_{i = 1}^{N} s_i = P = 4 r_{\text{max}} \int\limits_{0}^{\pi/2} \sqrt{1 - \tilde{e}^2 \sin{\theta}} \, d\theta, \\
r_{\text{max}} = \text{max}\{\chi, \varkappa\}, \\
r_{\text{min}} = \text{min}\{\chi, \varkappa\}, \\
\tilde{e} = \sqrt{1 - \left(\frac{r_{\text{min}}}{r_{\text{max}}}\right)^2},
\end{gathered}
\end{equation}
where $r_{\text{max}}$ and $r_{\text{min}}$ are semi-major and semi-minor axes, respectively, $\tilde{e}$ is an eccentricity, $P$ is a perimeter of the ellipse.  

\begin{figure*}[t]
  \centering
  
  \begin{subfigure}[t]{0.48\textwidth}
    \centering
    \includegraphics[width=\linewidth]{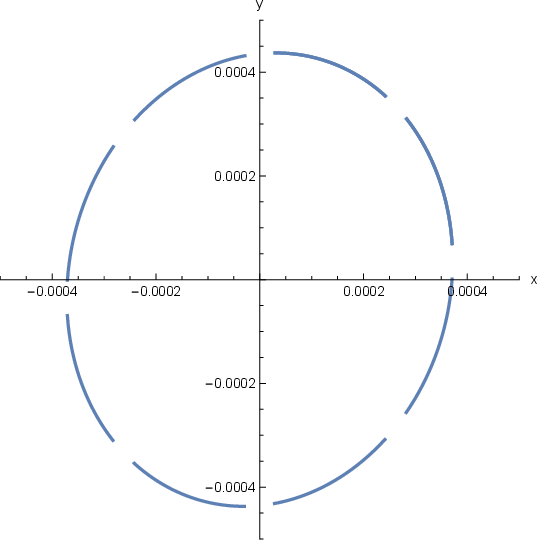}
    \subcaption{$T_{on} = 0.1 T$, $N=16$, $\phi_0 = 0$.}
    \label{fig:RMS_N=16_dt1=0.1T}
  \end{subfigure}
  \hfill
  \begin{subfigure}[t]{0.48\textwidth}
    \centering
    \includegraphics[width=\linewidth]{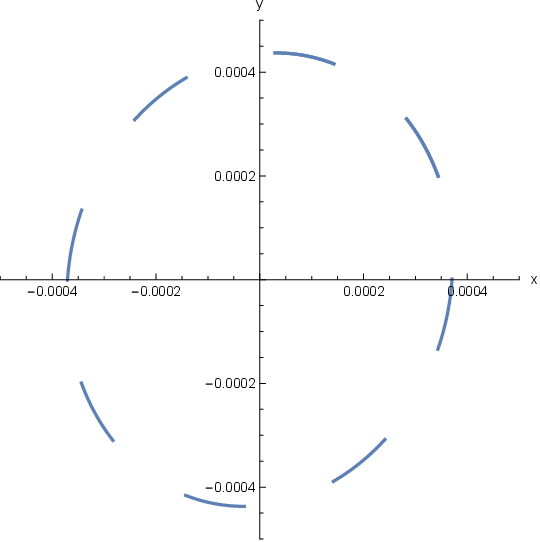}
    \subcaption{$T_{on} = 0.05 T$, $N=16$, $\phi_0 = 0$.}
    \label{fig:RMS_N=16_dt1=0.05T}
  \end{subfigure} 
  
  \caption{\emph{On and off} single-electron pencil beam's trajectory on the screen in the case of \emph{a capacitor model} [\emph{paraxial approximation}] (red) and parameters $\omega/\omega_c = 1$, $n=1/2$ and according to Table~\ref{Table:parameters}. See also Subfig.~\ref{fig:Capacitor_vs_textbook_k=1_n=0.5} for comparison.}
  \label{FIG:RMS}
\end{figure*}

For the illustration of partial arcs in the case of \emph{on and off} single-particle pencil beam, see Fig.~\ref{FIG:RMS}.

\begin{widetext}
Length measures of time expressed by the incomplete elliptic integral of the second kind (see Eq.~\eqref{lengthinellipse}) and dependent on unknown parameter $0 \leq \tilde{\alpha} \leq 1$, which is dimensionless time, follow
\bse
\begin{gather}
s_{on} = \sqrt{\frac{2}{N} (A_0 + D_0) \sum\limits_{i=1}^{N/2} \left(E\left[i \omega T_{on} + (i-1) \omega T_{off} + \phi_0 + \frac{\beta_0}{2}, m\right] - E\left[(i-1)\omega T_{on} + (i-1) \omega T_{off} + \phi_0 + \frac{\beta_0}{2}, m\right]\right)^2},  \label{EqRMSmeasure} \\
l_{on} = \frac{2}{N} \sqrt{A_0 + D_0}\sum\limits_{i = 1}^{N/2} \left(E\left[i \omega T_{on} + (i-1) \omega T_{off} + \phi_0 + \frac{\beta_0}{2}, m\right] - E\left[(i-1)\omega T_{on} + (i-1) \omega T_{off} + \phi_0 + \frac{\beta_0}{2}, m\right]\right), \label{EqAveragemeasure} \\
T_{on} = \tilde{\alpha} T, \quad
T_{off} = \frac{2}{N} T - T_{on} = \left(\frac{2}{N} - \tilde{\alpha} \right) T. 
\end{gather}
\ese
\end{widetext}

Let's assume the experimental values of length measures of time \eqref{EqRMSmeasure} and \eqref{EqAveragemeasure} are given as $s^*_{on}$ and $l^*_{on}$. A random search algorithm of minimization of a merit function $(s_{on} - s^*_{on})^2$ in the case of RMS measure (or $(l_{on} - l^*_{on})^2$ in the case of average measure), will give a solution for $\tilde{\alpha}$, which let's denote $\tilde{\alpha}^*$. For example, in Fig.~\ref{FIG:RMS}, $s^*_{on} \approx 0.2549$ mm and $l^*_{on} \approx 0.2544$ mm, and the random search algorithm is giving the value $\tilde{\alpha}^* = 0.1$, i.e., \emph{on beam} time $T_{on} = 200 $ ps.

%%%%%%%%%%%%%%%%%%%%%%%%%%%%%%%%%%%%%%%%%%%%%%%%%%%%%%%%%%%%%%%%%%%%%%%%%%%%%%%%%%%%%%%%%%%%%%%%%%%%%%%%%%%%%%%%%%%%%%%%%%%%%%%%%%%%%%%%%
%%%%%%%%%%%%%%%%%%%%%%%%%%%%%%%%%%%%%%%%%%%%%%%%%%%%%%%%%%%%%%%%%%%%%%%%%%%%%%%%%%%%%%%%%%%%%%%%%%%%%%%%%%%%%%%%%%%%%%%%%%%%%%%%%%%%%%%%%
\section{On the possibility of measuring the absolute arrival time of the first charged particle\label{Sec:arrival}}
%%%%%%%%%%%%%%%%%%%%%%%%%%%%%%%%%%%%%%%%%%%%%%%%%%%%%%%%%%%%%%%%%%%%%%%%%%%%%%%%%%%%%%%%%%%%%%%%%%%%%%%%%%%%%%%%%%%%%%%%%%%%%%%%%%%%%%%%%
%%%%%%%%%%%%%%%%%%%%%%%%%%%%%%%%%%%%%%%%%%%%%%%%%%%%%%%%%%%%%%%%%%%%%%%%%%%%%%%%%%%%%%%%%%%%%%%%%%%%%%%%%%%%%%%%%%%%%%%%%%%%%%%%%%%%%%%%%

Although experimentalists often assume that this device can measure only relative arrival times, I argue that this is not the case. In fact, the device can measure the arrival time of the first charged particle through the initial phase $\phi_0$ of the reference field, appearing in the phase relation
\be
\phi = \omega \Delta t + \phi_0, 
\ee
where $\Delta t$ denotes the relative time delay between two charged particles --- the first particle and any subsequent one --- and $\phi_0$ is the initial phase of the reference field at the moment the first particle enters the helical deflector. Experimentally, this phase is equivalent to the time at which the voltage is switched on. Therefore, from an experimental standpoint, the main challenge lies in accurately fixing the times at which the first electron is released and at which the voltage is applied.

%%%%%%%%%%%%%%%%%%%%%%%%%%%%%%%%%%%%%%%%%%%%%%%%%%%%%%%%%%%%%%%%%%%%%%%%%%%%%%%%%%%%%%%%%%%%%%%%%%%%%%%%%%%%%%%%%%%%%%%%%%%%%%%%%%%%%%%%%
%%%%%%%%%%%%%%%%%%%%%%%%%%%%%%%%%%%%%%%%%%%%%%%%%%%%%%%%%%%%%%%%%%%%%%%%%%%%%%%%%%%%%%%%%%%%%%%%%%%%%%%%%%%%%%%%%%%%%%%%%%%%%%%%%%%%%%%%%
\section{Deflection sensitivity\label{Sec:DS}}
%%%%%%%%%%%%%%%%%%%%%%%%%%%%%%%%%%%%%%%%%%%%%%%%%%%%%%%%%%%%%%%%%%%%%%%%%%%%%%%%%%%%%%%%%%%%%%%%%%%%%%%%%%%%%%%%%%%%%%%%%%%%%%%%%%%%%%%%%
%%%%%%%%%%%%%%%%%%%%%%%%%%%%%%%%%%%%%%%%%%%%%%%%%%%%%%%%%%%%%%%%%%%%%%%%%%%%%%%%%%%%%%%%%%%%%%%%%%%%%%%%%%%%%%%%%%%%%%%%%%%%%%%%%%%%%%%%%

Instantaneous dynamic deflection sensitivity, in the case of the ellipse, ranges in $[S_{\text{min}} (\omega), S_{\text{max}} (\omega)]$, has  
\begin{enumerate}
\item its maximum as 
\be
S_{\text{max}} (\omega) = \frac{r_{\text{max}} (\omega)}{U_0 \sin{(\omega t)}},
\ee
\item its minimum as 
\be
S_{\text{min}} (\omega) = \frac{r_{\text{min}} (\omega)}{U_0 \sin{(\omega t)}},
\ee
\item and its mean as
\be
\langle S \rangle = \frac{\sqrt{\chi (\omega) \kappa (\omega)}}{U_0 \sin{(\omega t)}},
\ee
the latter formula is taken according to the mapping of the ellipse to the \emph{phantom} circle using the equivalence of their areas $f_{e} = \pi \chi \varkappa$ and $f = \pi \langle r \rangle^2$, where the radius of that averaged circle is actually a mean deflection $\langle r \rangle = \sqrt{\chi \varkappa}$, representing a geometric mean.
\end{enumerate}

Since the instantaneous dynamic deflection sensitivity is a measure, which describes the apparatus' sensitivity of deflection to the instantaneous voltage, i.e., at the particular moment, the instantaneous voltage $U_0 \sin{(\omega t)}$ is used. For the comparison with other deflection systems, the deflection sensitivity with $U_0$ without $\sin{(\omega t)}$ at the denominator can be used.
\section{Comments and conclusions\label{Sec:concl}}
%%%%%%%%%%%%%%%%%%%%%%%%%%%%%%%%%%%%%%%%%%%%%%%%%%%%%%%%%%%%%%%%%%%%%%%%%%%%%%%%%%%%%%%%%%%%%%%%%%%%%%%%%%%%%%%%%%%%%%%%%%%%%%%%%%%%%%%%%
%%%%%%%%%%%%%%%%%%%%%%%%%%%%%%%%%%%%%%%%%%%%%%%%%%%%%%%%%%%%%%%%%%%%%%%%%%%%%%%%%%%%%%%%%%%%%%%%%%%%%%%%%%%%%%%%%%%%%%%%%%%%%%%%%%%%%%%%%

%========================
The present article demonstrates that \emph{the textbook model} \cite{ZhigarevBook, Gevorgian2015} (and the analogous magnetic-field model \cite{Varfolomeev1980}) for the helical deflector is actually an approximation of the newly introduced \emph{capacitor model} (with a similar magnetic analogue \cite{Smythe1989}). In turn, this \emph{capacitor model} should itself be regarded as an approximation \cite{Jefimenko1992, Ton1991} to the full electromagnetic theory obtained directly from Maxwell’s equations \cite{Buchholz1957, Gevorgian2022}.
Furthermore, the capacitor model may deviate from the real electromagnetic behavior because the helical system possesses both capacitance and inductance, whose effective values are frequency dependent in a slow-wave structure. As a result, at a characteristic resonant frequency \cite{Gevorgianidea}
\be
\omega_{LC} = \frac{1}{\sqrt{LC}}
\ee
the interaction becomes strongest, and the resulting electric field $\tilde{E}_\perp$ --- and hence the deflection amplitude ($\chi, \varkappa$) --- achieves its maximum. 

In helical slow-wave structures \cite{Basu1996, Basu1979, Rowe1965, Carter2018}, $\vb*{E}$ and $\vb*{B}$ fields are typically derived from the voltage and current on the transmission line, with the primary goal of analyzing wave propagation along the helix (e.g., phase velocity, dispersion, and impedance). The classical helix slow-wave models used differ from the present case. 

In \cite{Gevorgian2022}, the important problem of deriving the magnetic field of a two-wire helical line from the Poisson equation --- and thus directly from Maxwell’s equations --- was addressed, although only for the constant-current case. A comparable derivation for ribbons or other helical structures --- specifically, the determination of the electric field and the full electromagnetic field generated under an applied periodic voltage as a direct solution of Maxwell’s equations in full 3D geometry --- remains an open problem. Likewise, the resulting deflection characteristics and the validity of \emph{the textbook model} \cite{ZhigarevBook, Gevorgian2015} have not yet been fully established. The present article introduces a universal method for deriving the resulting deflection characteristics and addresses questions regarding the validity of \emph{the textbook model} \cite{ZhigarevBook, Gevorgian2015} by proposing a new \emph{capacitor model} and their comparison. It establishes a pathway from modeling to theory for high-precision instruments, such as the Advanced Picosecond Precision Radio Frequency Timer \cite{Zhamkochyan2024}. Moreover, the article \cite{Zhamkochyan2026} presents and experimentally validates a two-frequency RF helical-deflector system that converts the arrival time of keV electrons into a controlled spiral position on a detector, extending the temporal dynamic range while maintaining few-picosecond resolution.

%%%%%%%%%%%%%%%%%%%%%%%%%%%%%%%%%%%%%%%%%%%%%%%%%%%%%%%%%%%%%%%%%%%%%%%%%%%%%%%%%%%%%%%%%%%%%%%%%%%%%%%%%%%%%%%%%%%%%%%%%%%%%%%%%%%%%%%%%%%%%%%%%%%%%%%%%%%%%%%%

\acknowledgments
HLG acknowledges support from the Higher Education and Science Committee of Armenia in the frames of the research project 21AG-1C038 on \textit{Methods of Information Theory in Statistical Physics and Data Science}. HLG would like to express sincere gratitude to Prof. Armen Allahverdyan for valuable, fruitful, and engaging discussions.

%HLG would like to express his heartfelt gratitude to Anahit Shamamian for her invaluable guidance in directing him to the review by Varfolomeev, which led him to the paper on helical undulators and ultimately to Smythe's Book. Special thanks also go to Lekdar Gevorgian for his crucial assistance in locating the capacitance of a two-wire line in the Handbook of Physics, which significantly contributed to completing this work.

%\newpage
\appendix

\section{Parameters}

Here we present Table~\ref{Table:parameters}.
%T%T%T%T%T%T%T%T%T%T%T%T%T%T%T%T%T%T%T%T%T%T%T%T%T%T%T%T%T%T%T%T%T%T%T%T%T%T%T%T%T%T%T%T%T
\begin{table*}[h!]
    \centering
    \caption{Parameters of the helical deflector}\label{Table:parameters}
    \begin{tabular}{|c|c|c|c|c|c|c|c|c|c|c|c|c|c|c|} 
        \hline
        \hline
        $U_a$ & $U_0$ & $v_z /c$ & $z_0$ & $R$ & $d$ & $b$ & $\lambda$ & $\epsilon_d$ & $l$ & $\kappa = \lambda / (\pi d)$ & $\varphi_1, \varphi_2$ & $n$ & $\omega_c$ ($\nu_c$) & $L_D$ \\ 
        \hline
        2.5 keV & 10 V & 1/10 & $v_z t$ & $d/2$ & 10 mm & 1.5 mm & 60 mm & 1 & $\lambda n$ & $6/\pi$ &  $0, 2 n\pi$ & 1 & $\pi$ GHz (500 MHz) &  * \\
        \hline
        \hline
    \end{tabular}
\end{table*}
%T%T%T%T%T%T%T%T%T%T%T%T%T%T%T%T%T%T%T%T%T%T%T%T%T%T%T%T%T%T%T%T%T%T%T%T%T%T%T%T%T%T%T%T%T

\section{Magnetic field}\label{Appendix:B}
An open circuit offers infinite resistance to the flow of current, hence, the amount of current propagating through deflector is due to the time-dependent charge. So, the current in a single helix is equal
\be
\begin{gathered}
I(t) = \frac{d q(t)}{dt} = - \int\limits_{0}^{L} \frac{d \rho(t)}{dt} ds = - \frac{I_0}{2} \cos{(\omega t + \phi)} \overset{\Delta}{=} - \frac{I_0 (t)}{2}, 
\end{gathered}
\ee
where, for the parameters \ref{Table:parameters}, $I_0 = 2 \omega C U_0 L = 0.0108608$ A, and $L=6.7$ cm for $l=6 cm$ ($\lambda = 6$ cm and $n=1$). The Biot-Savart law gives the two-wire magnetic field
\be
\begin{gathered}
\vb*{B} (t) = \vb*{B}_0 \cos{(\omega t + \phi)}, \\
\vb*{B}_0 = - \frac{\mu_0}{4\pi} \int \frac{I_0 / 2 (d\vb*{s}_1 \times \vb*{r}_1 - d\vb*{s}_2 \times \vb*{r}_2)}{r^3} = \{B_{0x}, B_{0y}, B_{0z}\},
\end{gathered}
\ee
with contributions from the two spirals with the off-axis components of magnetic fields $I_1 (t) = - I_0 /2 \cos{(\omega t + \phi)}$, $r_{x1} = x_0 - x$, $r_{y1} = y_0 - y$, $r_{z1} = z_0 - z$ and $I_2 (t) = I_0 /2 \cos{(\omega t + \phi)}$, $r_{x2} = x_0 + x$, $r_{y2} = y_0 + y$, $r_{z2} = z_0 - z$, where the charged particle's off-axis distance is $\sqrt{x_0^2 + y_0^2}$. Hence, in non-paraxial case we have
\begin{widetext}
\be
\begin{gathered}
B_{0x} = \frac{\mu_0}{4\pi} \left(\int\limits_{\varphi_1}^{\varphi_2} \frac{\frac{I_0}{2R} \left[(\kappa y_0 - z_0 \cos{\varphi})/R + \kappa (\varphi \cos{\varphi} - \sin{\varphi}) \right] \, d\varphi}{S_1(x_0, y_0, z_0, \varphi)} + \int\limits_{\varphi_1}^{\varphi_2} \frac{\frac{I_0}{2R} \left[(\kappa y_0 - z_0 \cos{\varphi})/R + \kappa (\varphi \cos{\varphi} + \sin{\varphi}) \right] \, d\varphi}{S_2(x_0, y_0, z_0, \varphi)}\right), \\ 
B_{0y} = \frac{\mu_0}{4\pi} \left( \int\limits_{\varphi_1}^{\varphi_2} \frac{\frac{I_0}{2R} \left[ - (\kappa x_0 + z_0 \sin{\varphi})/R + \kappa (\cos{\varphi} + \varphi \sin{\varphi}) \right] \, d\varphi}{S_1(x_0, y_0, z_0, \varphi)} + 
\int\limits_{\varphi_1}^{\varphi_2} \frac{\frac{I_0}{2R} \left[ - (\kappa x_0 + z_0 \sin{\varphi})/R + \kappa (- \cos{\varphi} + \varphi \sin{\varphi}) \right] \, d\varphi}{S_2(x_0, y_0, z_0, \varphi)} \right), \\
B_{0z} = \frac{\mu_0}{4\pi} \left( \int\limits_{\varphi_1}^{\varphi_2} \frac{\frac{I_0}{2R} \left[ (x_0 \cos{\varphi} + y_0 \sin{\varphi})/R - 1 \right] \, d\varphi}{S_1(x_0, y_0, z_0, \varphi)} +  \int\limits_{\varphi_1}^{\varphi_2} \frac{\frac{I_0}{2R} \left[ (x_0 \cos{\varphi} + y_0 \sin{\varphi})/R + 1 \right] \, d\varphi}{S_2(x_0, y_0, z_0, \varphi)}\right).
\end{gathered}
\ee

% \be
% \begin{gathered}
% B_{0x} = \frac{\mu_0}{4\pi} \int\limits_{\varphi_1}^{\varphi_2} \frac{I_0/2R \left[(\kappa y_0 - z_0 \cos{\varphi})/R + \kappa (\varphi \cos{\varphi} - \sin{\varphi}) \right] \, d\varphi}{S_1(x_0, y_0, z_0, \varphi)} + \\
% + \frac{\mu_0}{4\pi} \int\limits_{\varphi_1}^{\varphi_2} \frac{I_0/2R \left[(\kappa y_0 - z_0 \cos{\varphi})/R + \kappa (\varphi \cos{\varphi} + \sin{\varphi}) \right] \, d\varphi}{S_2(x_0, y_0, z_0, \varphi)}, \\ 
% B_{0y} = \frac{\mu_0}{4\pi} \int\limits_{\varphi_1}^{\varphi_2} \frac{I_0/2R \left[ - (\kappa x_0 + z_0 \sin{\varphi})/R + \kappa (\cos{\varphi} + \varphi \sin{\varphi}) \right] \, d\varphi}{S_1(x_0, y_0, z_0, \varphi)} + \\
% + \frac{\mu_0}{4\pi} \int\limits_{\varphi_1}^{\varphi_2} \frac{I_0/2R \left[ - (\kappa x_0 + z_0 \sin{\varphi})/R + \kappa (- \cos{\varphi} + \varphi \sin{\varphi}) \right] \, d\varphi}{S_2(x_0, y_0, z_0, \varphi)}, \\
% B_{0z} = \frac{\mu_0}{4\pi} \int\limits_{\varphi_1}^{\varphi_2} \frac{I_0/2R \left[ (x_0 \cos{\varphi} + y_0 \sin{\varphi})/R - 1 \right] \, d\varphi}{S_1(x_0, y_0, z_0, \varphi)} + \\
% + \frac{\mu_0}{4\pi} \int\limits_{\varphi_1}^{\varphi_2} \frac{I_0/2R \left[ (x_0 \cos{\varphi} + y_0 \sin{\varphi})/R + 1 \right] \, d\varphi}{S_2(x_0, y_0, z_0, \varphi)}.
% \end{gathered}
% \ee
\end{widetext}

Let's evaluate a magnetic field. It is approximately equal to $[B] = \tfrac{\mu_0}{4\pi} \tfrac{I_0}{R} \approx 2 \times 10^{-7}$ T. His contribution to the Lorentz force will be $\tfrac{c}{10} [B] \approx 6$ V/m, much smaller than contribution of an electric field $[E] = Q_0 /R = 1242.83$ V/m.

In the paraxial approximation, the equations are  simplified as 
\be\label{magneticfield}
\begin{gathered}
\vb*{B}_\perp (t) = \frac{\mu_0}{4\pi} \int\limits_{\varphi_1}^{\varphi_2} \frac{I_0(t) \left[\kappa \varphi - z_0/ R \right]}{R} \left\{\begin{array}{c}  \cos{\varphi} \\ \sin{\varphi}
\end{array}\right\} \frac{d\varphi}{S(z_0, \varphi)}, \\
S(z_0, \varphi) = \left(1 + \left(z_0/R - \kappa \varphi \right)^2 \right)^{3/2}, \\
B_z = 0.
\end{gathered}
\ee

%%%%%%%%%%%%%%%%%%%%%%%%%%%%%%%%%%%%%%%%%%%%%%%%%%%%%%%%%%%%%%%%%%%%%%%%%%%%%%%%%%%%%%%%%%%%%%%%%%%%%%%%%%%%%%%%%%%%%%%%%%%%%%%%%%%%%%%%%%%%%%%%%%%%%%%%%%%%%%%%
%%%%%%%%%%%%%%%%%%%%%%%%%%%%%%%%%%%%%%%%%%%%%%%%%%%%%%%%%%%%%%%%%%%%%%%%%%%%%%%%%%%%%%%%%%%%%%%%%%%%%%%%%%%%%%%%%%%%%%%%%%%%%%%%%%%%%%%%%%%%%%%%%%%%%%%%%%%%%%%%
%%%%%%%%%%%%%%%%%%%%%%%%%%%%%%%%%%%%%%%%%%%%%%%%%%%%%%%%%%%%%%%%%%%%%%%%%%%%%%%%%%%%%%%%%%%%%%%%%%%%%%%%%%%%%%%%%%%%%%%%%%%%%%%%%%%%%%%%%%%%%%%%%%%%%%%%%%%%%%%%

%\begin{appendices}
%\end{appendices}

\end{document}